\documentclass[%
aps,
preprint,
nofootinbib,
eqsecnum,
a4paper,
showpacs,
 prd,
]{revtex4-2}
\usepackage{amsmath}
\usepackage{amssymb}
\usepackage{subfigure}
\usepackage{hyperref}
\usepackage[dvipsnames]{xcolor}
\usepackage{xcolor}
\usepackage{slashed}
\usepackage{mathrsfs}
\usepackage{braket}
\usepackage{graphicx}

\newcommand{\threeflavor}{{\begin{pmatrix}{\nu_e}\\ {\nu_{\mu}}\\ {\nu_{\tau}}\end{pmatrix}}}

\begin{document}
\title{Torsional modulation of atmospheric neutrino oscillation}

\author{Riya Barick}
\email{riyabarik7@gmail.com}
\author{Amitabha Lahiri}
\email{amitabha@bose.res.in}
\affiliation{S. N. Bose National Centre for Basic Sciences\\
	Block JD, Sector III, Salt Lake, WB 700106, INDIA.}

\begin{abstract}
Fermions act as sources of spacetime torsion. However, this torsion is non-dynamical and can be eliminated using its equations of motion. The resulting theory features an effective four-fermion interaction in a torsion-free background. Generically this interaction is non-universal and violates parity. When neutrinos propagate through matter, they experience the effect of this geometrical interaction, which is similar to the MSW effect, but diagonal in the mass basis. Since this quartic interaction term varies linearly with matter density, its effect will be more prominent for atmospheric neutrinos specially for upward going atmospheric neutrinos. We investigate the effect of spacetime geometry on $\nu_\mu \to \nu_\tau$ and $\nu_\mu \to \nu_e$ conversion probability and $\nu_\mu$ survival probability, by solving the Schr\"odinger equation using the {Preliminary Reference Earth Model (PREM)} density profile of the Earth. We also study the dependence of atmospheric neutrino oscillation probability on the CP phase angle in presence of this effect. We further provide a discussion of the MSW and parametric resonances in presence of the geometrical interaction. 
\end{abstract}
\maketitle
\newpage
\section{Introduction}
In 1998, the Super-Kamiokande neutrino observatory confirmed neutrino flavor oscillation~\cite{Super-Kamiokande}, 
using data from atmospheric neutrinos created by cosmic ray interactions in the Earth's atmosphere. Neutrino oscillations typically indicate non-zero neutrino masses that are non-degenerate among the flavors. Neutrino flavor oscillation is thus one of the most compelling signals of new physics beyond the Standard Model (BSM), as it is not expected in the context of the Standard Model (SM).

The conventional paradigm states that there are three flavor eigenstates, or weak gauge eigenstates, namely $\nu_e, \nu_\mu$ and $\nu_\tau$, which are superpositions of three mass eigenstates, namely $\nu_1, \nu_2$, and $\nu_3$ with the masses $m_1, m_2$, and $m_3$, respectively. The $3 \times 3$ mixing matrix for Dirac neutrinos is parametrized by three angles $\theta_{12}, \theta_{13}, \theta_{23} $ and one CP-violating phase $\delta_{CP}$\,, while two additional phases are needed if the neutrinos are  Majorana particles. Neutrino flavor oscillation probabilities are functions of these four parameters and the two independent mass squared differences: $\Delta m_{21}^2 = m_2^2-m_1^2$ and $\Delta m_{31}^2 = m_3^2-m_1^2 $ \cite{Mohapatra:1998rq}.

Atmospheric neutrinos are produced by the decays of pions and kaons generated from cosmic ray interactions with nucleons in the Earth’s atmosphere. These neutrinos exhibit a wide range of energies $(\simeq 0.1 $ GeV to $>$1 TeV) and travel substantial distances before detection $(\simeq 10$ km to $\simeq 10^4$ km). They encompass all flavors of neutrinos and antineutrinos.
By considering their primary production mechanisms, we can derive some generic relations for the flux ratios of different neutrino flavors without detailed calculations. For instance, from the decay chain of a charged pion $\pi^+ \rightarrow \mu^+ \nu_\mu$ followed by $\mu^+ \rightarrow e^+ \nu_e \bar{\nu}_\mu$ (and the
charge conjugated process for $\pi^-$), the ratio $\frac{\nu_\mu + \bar{\nu}_\mu}{\nu_e + \bar{\nu}_e}$ is expected
to be around 2 at energies $\simeq 1$ GeV, which is the typical energy scale of most muons decaying in the atmosphere \cite{ParticleDataGroup:2024cfk}. 
At higher energies, some muons reach the Earth before decaying, causing the ratio to increase.
Furthermore, the zenith angle distributions of atmospheric neutrinos are expected to be symmetric between upward going and downward going neutrinos. This holds true for energies above 1 GeV. However, at lower energies, the Earth’s geomagnetic field induces up-down asymmetries in the primary cosmic rays. The zenith angle corresponds to the flight length of atmospheric neutrinos. Vertically upward going neutrinos travel $\simeq 10^4$ km from the other side of the Earth, while downward going neutrinos, produced just above the experimental site, travel about 10 km before detection.

The atmospheric neutrino fluxes have been calculated in detail based on the energy spectrum and composition of primary cosmic rays and their hadronic interactions in the atmosphere. These calculations also consider the effects of solar activity and the geomagnetic field. Results from several groups are available~\cite{Honda1,Honda2}, with a typical uncertainty in the absolute flux of $10 - 20 \% $ , while the uncertainty in the ratio of fluxes between different flavors is much smaller $(<5 \%)$.
Several projects for atmospheric neutrino observations are either proposed or under preparation. The atmospheric neutrino observation program is included in plans for future neutrino telescopes such as ORCA in the KM3NeT project~\cite{KM3} in the Mediterranean Sea and the IceCube Upgrade~\cite{IceCube-Gen2:2020qha}. Additionally, future large underground detectors like Hyper-Kamiokande~\cite{Hyper-Kamiokande:2018ofw} and DUNE~\cite{DUNE:2020jqi,DUNE:2022aul} will also study atmospheric neutrinos.

Although the neutrino oscillation itself is evidence of BSM physics, the existence of new physics can
conversely modify the standard three-neutrino mixing paradigm, which leads to reinterpretations
of the oscillation data. {An example of such new physics is the torsion generated by fermions in curved spacetime, which is generally not taken into account in BSM physics. 
This torsion is a non-dynamical field, so it may be integrated out, resulting in an effective current-current interaction which can be non-universal and chiral.
Then the contribution of this torsional, or geometrical, four-fermion interaction on neutrino oscillations
may not be negligible. In this paper we investigate some possible effects of this novel interaction on the atmospheric oscillations of neutrinos.}

Other authors have also explored neutrino oscillation scenarios involving torsion~\cite{DeSabbata:1981ek, Alimohammadi:1998vx, Zhang:2000ue,Adak:2000tp, Zubkov:2013zxa, Fabbri:2015jaa, JulioCirilo-Lombardo:2019shi}. {However, there are some crucial differences between these and related works and our model, which we list here: \\
a) Unlike those works, we do not consider a fixed background torsion; instead, the torsion is generated dynamically by the fermions in the background;\\
b) The coupling of this torsion to fermions is taken to be chiral and non-universal;\\
c) The torsion field itself is non-propagating and thus can be integrated out from the action, leaving behind a four-fermion interaction term;\\
d) The coupling constants are not required to be of order unity, which allows the effect to be non-negligible.
 }
 
{The layout of the paper is as follows. In Sec.~\ref{gravity} we provide a brief outline of the origin of the geometrical four-fermion interaction from dynamically generated torsion in the first-order approach to gravity. Then in Sec.~\ref{3nu} we demonstrate how this geometrical, or torsional, quartic interaction contributes to the neutrino mass matrix. In Sec.~\ref{oscillograms}, we explore the effects of this interaction on the atmospheric $\nu_\mu$ conversion and survival probabilities by solving the modified neutrino propagation equation through the Earth, using the Preliminary Reference Earth Model (PREM) density profile. Sec.~\ref{resonance-derivation} gives a discussion of the MSW and parametric resonance conditions in presence of this torsional interaction. Finally, in Sec.~\ref{summary} we summarize the work presented in this paper and outline possible directions for future research.}

\section{Fermions and torsion}\label{gravity}
The presence of matter changes the geometry of spacetime, which in turn affects the dynamics of matter. 
If the curvature of the spacetime is not large enough, we can neglect the effect of spacetime geometry on the bosonic fields but the dynamics of fermions in a curved spacetime is different. Spacetime geometry gives rise to an effective four fermion interaction involving coupling constants which are different for different species and can only be fixed from experimental observations. From simple considerations, we can expect the interaction to be weaker than weak interactions, so it will be overwhelmed by charged current interactions in most processes. A notable class of exception is that of neutrino oscillations in matter. Let us first briefly discuss the origin of  the geometrical four-fermion interaction~\cite{Chakrabarty:2019cau,Barick:2023wxx}.

Usually the dynamics of fermions in curved spacetime is described in the Einstein-Cartan-Sciama-Kibble (ECSK) formalism, a first order formulation of gravity, in which the variables are tetrads and the spin connection~\cite{Cartan:1923zea,Cartan:1924yea,Kibble:1961ba,Sciama:1964wt,Hehl:1976kj,Hehl:1974cn,Hammond:2002rm,Hehl:2007bn,Poplawski:2009fb,Poplawski:2009fb,Mielke:2017nwt,Chakrabarty:2018ybk}. The tetrad field  $e^\mu_a$  connects the ``internal flat space'' (isomorphic to the tangent space at each point) metric $\eta_{ab}$ to spacetime metric $g_{\mu \nu}$ through the relations 
\begin{align}
	\eta_{{ab}}e^a_\mu e^b_\nu = g_{\mu\nu}\,, \quad g_{\mu\nu}e^\mu_a e^\nu_b = \eta_{ab}\,,
	 \quad e^\mu_a e^a_\nu = \delta^\mu_\nu\,,
\end{align}
where  $e^a_{\mu}$ is the inverse tetrad,  also called co-tetrad. {Its determinant is related to the metric determinant as $|e| = \sqrt{|g|}$\,.} 
{We will use Greek indices for the spacetime and Latin indices for the internal flat space.}
{Dirac matrices $\gamma^a$ are defined on the internal flat space as $\left[\gamma_a\,, \gamma_b\right]_+ = 2\eta_{ab}$\,, but we can also define spacetime $\gamma$-matrices by $\gamma^\mu := e^\mu_a \gamma^a\,,$ which satisfy $\left[\gamma_\mu\,, \gamma_\nu\right]_+ = 2g_{\mu\nu}\,.$}
 Metric-compatibility of the tetrad produces
\begin{align}
 \nabla_\mu e_\nu^a = 0 \quad
\Rightarrow \quad  \partial_\mu e^a_\nu + 	A_{\mu}{}^{a}{}_{b} e^b_\nu  - \Gamma^\lambda{}_{\mu \nu} e^a_\lambda = 0\,.
	\label{tetrad-postulate}
\end{align}
It is analogous to metric compatibility of the Levi-Civita connection $\widehat{\nabla}.$ In the presence of fermions,  {the connection need not be torsion-free}  and then the connection components $\Gamma^\lambda_{\mu \nu}$ are not symmetric. The antisymmetric part of the connection $\Gamma^\lambda_{\mu \nu}$ is known as torsion 
\begin{align}\label{torsion-def}
C^\lambda_{\mu \nu} = \Gamma^\lambda_{\mu \nu}-\Gamma^\lambda_{\nu \mu}.
\end{align}
In Eq.~(\ref{tetrad-postulate}) $A_{\mu}{}^{a}{}_{b}$ is called spin connection and it appears in the covariant derivative of spinors,
\begin{equation}\label{Dirac-operator}
	D_\mu\psi = \partial_\mu\psi -\frac{i}{4} A_\mu{}^{ab} \sigma_{ab}\psi\,, \qquad \sigma_{ab} = \frac{i}{2}\left[\gamma_a\,, \gamma_b\right]_{-}\,.
\end{equation}
{We will be using the $(+---)$ signature in this paper.}
The spin connection $A_{\mu}{}^{ab}$ has two parts --- the torsion-free Levi-Civita spin connection $\omega_{\mu}{}^{ab}$ and the contorsion $\Lambda_{\mu}{}^{ab}$, 
\begin{equation}\label{split}
	A_{\mu}{}^{ab} = \omega_{\mu}{}^{ab}  + \Lambda_{\mu}{}^{ab}\,.
\end{equation}
If we replace $A_{\mu}{}^{ab}$ in Eq.~(\ref{tetrad-postulate}) by $\omega_{\mu}{}^{ab}$, the affine connection $\Gamma^\lambda_{\mu \nu}$ becomes the Christoffel symbol $\widehat{\Gamma}^\lambda_{\mu \nu}$. Similarly if the contorsion $\Lambda_{\mu}{}^{ab}$ is contracted with tetrads in Eq.~(\ref{tetrad-postulate}), we get the antisymmetric part of the connection also known as torsion as given in Eq.~(\ref{torsion-def}). 

In terms of the tetrads and the spin connection, the Ricci scalar can be written as
\begin{align}\label{Ricci}
	R(\Gamma) &= F_{\mu\nu}{}^{ab} e^\mu_a e^\nu_b\,, 
 \end{align}
where $F_{\mu\nu}{}^{ab}$ is the field strength of the spin connection, 
\begin{align}\label{Field strength}
F_{\mu\nu}{}^{ab} &= \partial_\mu A_\nu{}^{ab} - \partial_\nu A_\mu{}^{ab} + A_{\mu}{}^{a}{}_{c} A_{\nu}{}^{cb} -  A_{\nu}{}^{a}{}_{c} A_{\mu}{}^{cb}\,.
\end{align}
The total action of gravity with a minimally coupled fermion field is written as
\begin{align}\label{action.1}
&{S = S_{tetrad}[e] + \int |e| d^4x \left[ \frac{i}{2} \left(\bar{\psi}\slashed{D}\psi - \left(\bar{\psi}\slashed{D} \psi \right)^\dagger  \right) - m \bar{\psi} \psi \right] \,,} 
\end{align}
where $S_{tetrad}[e]$ is the action due to gravity in tetrad formalism and is given by
\begin{align}\label{gravity-action}
S_{tetrad}[e] = \frac{1}{2\kappa} \int |e| d^4x F_{\mu\nu}{}^{ab} e^\mu_a e^\nu_b \,, \qquad \mathrm{with} \qquad \kappa = 8\pi G\,,
\end{align}
and the derivative $D_\mu$ acts on the spinor via spin connection as given in Eq.~(\ref{Dirac-operator}).

 {We can write} the Lagrangian density of the fermion field as 
\begin{align}
	{\mathscr L}_\psi &= \frac{i}{2}	\left(\bar{\psi}\gamma^\mu\partial_\mu\psi - 
	\partial_\mu\bar{\psi}\gamma^\mu\psi - \frac{i}{4} A_{\mu}{}^{ab} \,
	\bar{\psi}[\sigma_{ab}, \gamma_c ]_{_+} \psi\, e^{\mu c} \right) 
    -m\bar\psi\psi\,.\label{L_psi}
\end{align}
Using the split connection as given in Eq.~(\ref{split}), the total action can be written as 
\begin{align}
	S =&  \frac{1}{2\kappa}\int |e| d^4x\, e^\mu_a e^\nu_b\left({F[\omega]_{\mu\nu}{}^{ab} }+   \partial_{[\mu}\Lambda_{\nu]}{}^{ab} 
	+	\left[\omega_{[\mu}, \Lambda_{\nu]}{}^{ab}\right]_{-}\right) \notag \\
	 & \qquad  + \int |e| d^4x\, 
	\left(	\frac{1}{2\kappa}  e^\mu_a e^\nu_b\left[\Lambda_{[\mu }, \Lambda_{\nu]}{}^{ab}\right]_{-}  
	+	{\mathscr L}_\psi \right)\,.
\label{action}
\end{align}
{The last two terms on the first line of Eq.~(\ref{action}) add up to a total derivative and thus give no contribution when we } vary the action with respect to the contorsion $\Lambda$\,. {This is distinct from theories where (con)torsion is a dynamical field. For torsion  to be dynamical, the theory must have additional terms involving derivatives of torsion.} From the second line we get {an algebraic equation for the contorsion,}
\begin{align}\label{spin-connection}
 {\Lambda_\mu{}^{ab} = \frac{\kappa}{4}\epsilon_{abcd}e^c_\mu\,\bar{\psi}\gamma^d\gamma^5 \psi \,,}
\end{align}
where we have used  $[\gamma_c,\sigma_{ab} ]_{_+} = 2\epsilon_{abcd}\gamma^d\gamma^5 $\,. Since this is a non-dynamical solution we can put back the above expression of $\Lambda_\mu{}^{ab}$ into the action,
{to find that} in curved spacetime, the Dirac equation becomes nonlinear~\cite{Gursey:1957,Finkelstein:1960,Hehl:1971qi, Gasperini:2013},
\begin{equation}\label{DE1}
	i\slashed{\partial} \psi + \frac{1}{4}\omega_\mu{}^{ab} \gamma_\mu \sigma_{ab} \psi
	- m\psi 		- \frac{3\kappa}{8}\left(\bar{\psi}\gamma^a\gamma^5\psi\right)\gamma_a\gamma^5\psi = 0\,.
\end{equation}
{We also get the Einstein equation for a torsion-free connection but with a four-fermion interaction term in the energy-momentum tensor.}
 {So far we have worked with} a single fermionic field, but of course we must include all fermions in the action. Generally, torsion will couple to the left-handed component and right-handed component of a fermionic field differently with different coupling constants. We can write the most general form of the fermion Lagrangian as
\begin{align}
	{\mathscr L}_\psi &= \sum\limits_{i}	\left(\frac{i}{2}\bar{\psi}_i\gamma^\mu\partial_\mu\psi_i - 
	\frac{i}{2}\partial_\mu\bar{\psi}_i\gamma^\mu\psi_i 	+ \frac{1}{8} \omega_{\mu}{}^{ab} e^{\mu c}  \,
	\bar{\psi}_i[\sigma_{ab}, \gamma_c ]_{_+} \psi_i\, -m\bar\psi_i\psi_i\, \right. \notag \\ 
	&\qquad \qquad \left.
	+ \frac{1}{8} \Lambda_{\mu}{}^{ab} e^{\mu c}
	\left(\lambda^i_{L}\bar{\psi}_{iL} \left[\gamma_c, \sigma_{ab}\right]_{+}\psi_{iL} + \lambda^i_{R}\bar{\psi}_{iR} \left[\gamma_c, \sigma_{ab}\right]_{+}\psi_{iR}\right)
	\right)\,,\label{L_psi_all}
\end{align}
where the sum includes all fermion species. After including this Lagrangian in the action of Eq.~(\ref{action}) and varying with respect to $\Lambda\,,$ we again find an algebraic expression for the contorsion, which is now a chiral object,
\begin{equation}\label{chiral.torsion}
	\Lambda_{\mu}{}^{ab} = \frac{\kappa}{4}\epsilon^{abcd}e_{c\mu} \sum\limits_i \left(-\lambda^i_{L}\bar{\psi}_{iL}\gamma_d \psi_{iL} + \lambda^i_{R}\bar{\psi}_{iR}\gamma_d \psi_{iR}\right)\,.
\end{equation}
As before, we can insert this solution back into the action and get the effective four-fermion interaction term,
\begin{equation}\label{4fermi}
	-\frac{3\kappa}{16}\left(\sum\limits_i \left(-\lambda^i_{L}\bar{\psi}^i_{L} \gamma_a \psi^i_L + \lambda^i_{R}\bar{\psi}^i_{R} \gamma_a \psi^i_{R}\right)\right)^2\,.
\end{equation}
We will refer to this four-fermion term as the geometrical interaction or as the torsional interaction term. This four-fermion interaction can be rewritten using the vector and axial currents as
\begin{equation}\label{V-A}
	-\frac{1}{2}\left(\sum\limits_i \left(\lambda_i^V \bar{\psi}^i \gamma_a \psi^i + \lambda_i^A \bar{\psi}^i \gamma_a\gamma^5 \psi^i\right)\right)^2\,,
\end{equation}
where we have written $\lambda^{V, A} = \frac{1}{2}(\lambda_R \pm \lambda_L)$\, and absorbed a factor of $\sqrt{\frac{3\kappa}{8}}$\, in $\lambda_{V, A}\,.$ This geometrical quartic interaction term will contribute to the effective masses of fermions passing through a background of fermionic matter.

{A few things should be mentioned here. One is that the interaction does not vanish in the ``flat space limit'' $\omega_\mu{}^{ab} \to 0$\,. This is not at odds with the fact that the interaction is an effect of spacetime being curved, since spacetime is only approximately flat in the presence of matter. Thus in terrestrial experiments, where $\omega_\mu{}^{ab}$ can be neglected and the spacetime is approximately flat, the geometrical interaction does not disappear but can be tested in experiments. The Ricci tensor or other curvature-related quantities do not affect the interaction --- after eliminating $\Lambda$ from the action using Eq.~(\ref{chiral.torsion}), what we are left with is a theory of ordinary torsion-free general relativity with fermions, but with a new chiral, non-universal, four-fermion interaction.
Another important point is that while a four-fermion interaction is not renormalizable in flat space, renormalizability is not an applicable yadstick in this case, because spacetime will not remain flat if we consider loop integrations with sufficiently high cutoff. Since the interaction arises from the dynamics of fermions on curved spacetime, it is not possible to ignore the effect of gravity when it does become important.}
{Furthermore, as we have absorbed a factor of $\sqrt{\frac{3\kappa}{8}}\,$ into  $\lambda\,$ in Eq.~(\ref{V-A}), one might think that the interaction is ``suppressed by'' 
a factor of $\sqrt{\kappa} \sim M_P^{-1}.$ But that would be an incorrect conclusion. The contorsion field $\Lambda_\mu^{ab}$ is not a quantum field, it is not even a propagating field, so its couplings with the fermions currents are not required to be small and thus $\sqrt{\kappa}$ is not the scale of the $\lambda$. The absorbed factor only sets the mass dimension of $\lambda$ to be $-1$, we cannot determine the size of the $\lambda$'s from theoretical considerations alone, each $\lambda$ has to be fixed from experimental observations. However, the interaction term should be small compared to the free Hamiltonian since we will use perturbation theory. We can thus expect this interaction to be no larger than the effective quartic interaction we obtain at low energies from weak interactions, perhaps somewhat smaller to avoid conflict with known experimental results. Also, there is no such symmetry which forces $\lambda\,$ to be zero, so we must consider the effect of the geometrical four-fermion interaction in our calculations.}

\section{Neutrino oscillation due to torsional interaction}\label{3nu}
In this paper, we are interested in investigating the effect of the geometrical four-fermion interaction on the conversion and survival probabilities of atmospheric neutrinos  while propagating through the Earth. {In addition to the usual matter interaction term arising from weak interactions,} the geometrical interaction Lagrangian of neutrinos with the background is given by
\begin{equation}
\label{bkgd.int}
		-\left(\sum_{i=1,2,3}\left(\lambda_{i}^{V}\bar{\nu}_i \gamma_a \nu_i + \lambda_{i}^A \bar{\nu}_i \gamma_a\gamma^5 \nu_i \right)\right) \times
			\left(\sum_{f=e, p, n}\left(\lambda_{f}^{V}\bar{f} \gamma_a f + \lambda_{f}^A \bar{f} \gamma_a\gamma^5 f \right)\right) \,.
\end{equation}
Here $\nu_i$ is the neutrino field propagating through the Earth and $f$ runs over the fermionic fields present in the background matter. The Earth matter background consists of electrons, protons and neutrons (or electrons, up and down quarks). Also in the second factor of the above expression we have neglected self-interaction of neutrinos since their density in normal matter is several orders of magnitude lower than that of electrons or hadrons.
{For the Earth,} typical matter densities are not sufficient to induce significant curvature, thus the calculations can be performed on a flat background.

Similarly to weak interactions, the contribution from the background fermions can be approximated by its average value by considering the forward scattering of neutrinos~\cite{Wolfenstein:1977ue, Pal:1989xs, Ghose:2023ttq}. For a non-relativistic fermionic matter background, the average of the second factor in 
Eq.~(\ref{bkgd.int}) for fermions of type $f$ is the weighted number density 
%
\begin{align}
\tilde{n} = \sum_f \lambda_f n_f,
\label{weighted_no._density}
\end{align} 
where $ n_f$ is the number density of background fermions. Thus the quartic interaction term looks like
\begin{equation}
\label{quartic-int}
  -\left(\sum_{i=1,2,3}\left(\lambda_{i}^{V}\bar{\nu}_i \gamma_0 \nu_i + \lambda_{i}^A \bar{\nu}_i\gamma_0 \gamma^5 \nu_i \right)\right) \, \tilde{n}\,.
\end{equation}

Since only left-handed neutrinos are seen in nature, we can write 
the contribution to the effective Hamiltonian due to the torsional four-fermion interaction as
\begin{equation}
 \sum_{i=1,2,3}\left(\lambda_{i}{\nu}_i^\dagger \nu_i \right)\,\tilde{n}\,.
\end{equation}

Analytical expressions of the oscillation probabilities of
neutrinos, created from cosmic ray interactions in the atmosphere, for their propagation through the Earth, have been found using certain approximations~\cite{Gandhi:2004md, Gandhi:2004bj, Gandhi:2007td, Ohlsson:1999um,Freund:1999vc,Blennow:2013rca, Barger:1980tf,Zaglauer:1988gz,Kimura:2002hb, Barger:1980tf, Yasuda:1998mh, Freund:2001pn,Kimura:2002wd,Akhmedov:2004ny,Nunokawa:2005nx}. Let us therefore describe  what happens when the geometrical four-fermion interaction term is considered in addition to the usual matter interactions. The neutrino flavor eigenstates $\nu_{\alpha} $ are the linear combinations of the mass eigenstates $\nu_{i}$ defined as
\begin{equation}
	\ket{\nu_{\alpha}}=\sum_{i}U^{*}_{\alpha i}\ket{\nu_{i}}\,,
	\label{3.mixing}
\end{equation}
where $U$ is a unitary matrix, known as the Pontecorvo-Maki-Nakagawa-Sakata(PMNS) mixing matrix, defined by three mixing angles and a Dirac-type phase, as
\begin{align}
        U&=\begin{pmatrix}c_{12}c_{13} & s_{12}c_{13} & s_{13}e^{-i\delta_{CP}}\\-s_{12}c_{23}-c_{12}s_{23}s_{13}e^{i\delta_{CP}} & c_{12}c_{23}-s_{12}s_{23}s_{13}e^{i\delta_{CP}} & s_{23}c_{13}\\ s_{12}s_{23}-c_{12}c_{23}s_{13}e^{i\delta_{CP}} & -c_{12}s_{23}-s_{12}c_{23}s_{13}e^{i\delta_{CP}} & c_{23}c_{13}\end{pmatrix}\,,
\end{align}
where $c_{ij} = \cos \theta_{ij}$, $s_{ij} = \sin \theta_{ij}$ and $\delta_{CP}$ is the leptonic
CP-violating phase \cite{Kobayashi:1973fv,Cabibbo:1977nk}. In general, the angles $\theta_{ij}$ can be taken to lie in the first quadrant, whereas the CP-violation phase $\delta_{CP}$ is taken to
be between 0 and 2$\pi$~\cite{ParticleDataGroup:2024cfk}.

{The equation for neutrinos }passing through the Earth matter, including the effects of weak interaction as well as the geometrical four-fermion interaction, is given by 
\begin{align}
i\frac{d}{dt}\begin{pmatrix}{\nu_1}\\ {\nu_2} \\ \nu_3\end{pmatrix}=\left[E+\frac{1}{2E}\begin{pmatrix}m_1^2 & 0 & 0 \\ 0 & m_2^2 & 0 \\ 0 & 0 & m_3^2\end{pmatrix}+\begin{pmatrix}\lambda_1 & 0 & 0 \\ 0 & \lambda_2 & 0 \\ 0 & 0 & \lambda_3\end{pmatrix}\tilde{n}-\frac{G_F}{\sqrt{2}}n_n+U^{\dagger}\begin{pmatrix}A & 0 & 0 \\ 0 & 0 & 0 \\ 0 & 0 & 0\end{pmatrix}U\right]\begin{pmatrix}\nu_1 \\ \nu_2 \\ \nu_3 \end{pmatrix}\,, \label{eq:TDSE_for_3nu.a}
\end{align}
where {$A$ is the matter potential due to weak interaction only,} $A={\sqrt{2}}G_F n_e$\,.
{In the flavor basis we can write this as} 
\begin{align}
	i\frac{d}{dt}\threeflavor&=\left[E'_0\mathbb{I}+\frac{\Delta \tilde{m}^2_{31}}{2E}\left[U\begin{pmatrix}0 ~& 0 & 0 \\ 0 ~& \tilde{\alpha} & 0 \\ 0 ~& 0 & 1\end{pmatrix}U^{\dagger}+\begin{pmatrix}\tilde{A} & 0 & 0 \\ 0 & 0 & 0 \\ 0 & 0 & 0\end{pmatrix}\right]\right]\,\threeflavor \,,\label{eq:TDSE_for_3nu.b}
\end{align}
%
with $E'_0=E+\frac{m_1^2+2\lambda_1\tilde{n}E}{2E}-\frac{G_F}{\sqrt{2}}n_n\,.$  {The torsionally modified} mass-squared difference is defined as 
\begin{align}
\Delta \tilde{m}^2_{ij} := \Delta m^2_{ij} +2 \tilde{n} E \Delta\lambda_{ij}\,,
\label{torsion-induced-mass-splitting}
\end{align}
where $\tilde{n}=\sum_f \lambda_f n_f\,,$ as given in Eq.~(\ref{weighted_no._density}), {$\Delta \lambda_{ij} = \lambda_i - \lambda_j\,$ is the difference of geometrical coupling constants of neutrinos,} and $\tilde{\alpha}$ is defined as the ratio of two mass squared differences, $\tilde{\alpha} =\frac{ \Delta \tilde{m}_{21}^2}{\Delta \tilde{m}_{31}^2}\,$. We  {will assume an} electrically neutral matter background, $n_p=n_e\,,$ and  also that $n_p=n_n\,.$ 
For convenience, we will also assume that the geometrical couplings of the background fermions are approximately equal, $\lambda_e\simeq\lambda_p \simeq\lambda_n\,,$ so that $\tilde{n} = 3\lambda_e n_e$\,. {Furthermore, since we are considering simulations to test the effect of the new interaction, we will assume for simplicity that all the $\lambda_f$ and $\Delta \lambda_{ij}$ are of the same order of magnitude. In what follows, the product of $\lambda_f$ and $\Delta \lambda_{ij}$ will be represented as $\lambda^2\,$ for all the combinations $f,i,j$.} Of course, if we were to fit actual data, we would need to treat each of these combinations separately. 
We have defined the matter potential in presence of torsion as 
\begin{align}
\tilde{A} = \frac{2EA}{\Delta \tilde{m}^2_{31}} = \frac{2\sqrt{2}G_F n_e E}{\Delta \tilde{m}^2_{31}}\,.
\label{matter-potential-including-torsion}
\end{align}
%

%
%

We have solved Eq.~(\ref{eq:TDSE_for_3nu.b}) analytically for small $\tilde{\alpha}$ and $\sin{ \theta_{13}}$\,, making a perturbative expansion in those parameters and keeping terms up to second order in them.  {Analytical expressions for the probabilities of oscillations between different neutrino flavors were found in certain approximations
and given in~\cite{Barick:2023wxx,Barick:2023qjq}.} {For the sake of completeness, we have included them in Appendix~\ref{analytics}.
}


In this paper we investigate how the conversion and survival probabilities of atmospheric neutrinos are modified by the geometrical four-fermion interaction. We have solved Eq.~(\ref{eq:TDSE_for_3nu.b}), which we will call the Schr\"odinger equation for the propagation of neutrinos through the Earth, numerically using the Preliminary Reference Earth Model~\cite{Dziewonski:1981xy} (PREM) density profile.  {In the next section, we plot and analyze} the $ \nu_\mu \to \nu_\tau$ and $\nu_\mu \to \nu_e $ conversion probabilities, as well as the $ \nu_{\mu} $ survival probability, as functions of both energy and zenith angles. Such plots are usually called oscillograms~\cite{Ohlsson:1999um,Chizhov:1998ug,Akhmedov:2006hb,Kajita:2004ga}. In these plots we have used the following values of the neutrino oscillation parameters: {$\Delta m_{21}^2 = 7.41\times 10^{-5}~\text{eV}^2$, $\Delta m_{31}^2 = {{2.507\times 10^{-3}~\text{eV}^2}} ({{-2.4119\times 10^{-3}~\text{eV}^2}})\,, \theta_{23}=42.2^\circ\, (49.0^\circ), \theta_{12} = 33.41^\circ\,, \theta_{13} = 8.58^\circ\, (8.57^\circ),$ {and} $\delta_{CP}=232^\circ\, (276^\circ)\,.$} {These numbers are for normal hierarchy (NH) and those in parentheses are for inverted hierarchy (IH), and are best-fitted values} including the Super-Kamiokande atmospheric data~\cite{ParticleDataGroup:2024cfk}. 

{We note here that the analytical results given in Appendix~\ref{analytics} are not central to our analysis in this paper. 
Our main results are given as plots, which are obtained not from perturbative expansions but by solving Eq.~(3.8) on a
computer.  We rely on the analytical results as a guide to understand why the plots are the way 
they are. In particular, for the range of energies and matter densities appropriate to the atmospheric 
neutrinos, the parameter $\Tilde{\alpha}$ does not remain very small. }

{In our analysis, we have employed the
relation between the baseline, the maximum travel distance before detection, and the zenith angle (defined as the angle between the vertical direction and the incoming neutrino direction), in the context of upward-going neutrinos} 
\begin{align}
   L = -2R\cos \theta_\nu \,,
   \label{baseline-zenith}
\end{align}
where $L\,$ is the neutrino baseline, $R\,$ is the radius of the Earth, and $\theta_\nu$ is the zenith angle. 

\section{Effect of the torsional interaction on oscillograms \label{oscillograms} }

%
For our simulations, we have chosen the coupling constants $\lambda_i$ in the following way. In a recent work~\cite{Barick:2025ahc}, we used a simulation based on the DUNE experiment parameters to calculate bounds on the geometrical couplings using a $\chi^2$ analysis. The bound on $\lambda\Delta\lambda$ was found to be approximately $\pm 0.012~ G_F$\,, assuming background fermions to be $e,u,d$, with all of them having the same $\lambda$\,, so the total $\lambda\Delta\lambda$ comes to about $\pm 0.09~G_F$. On the other hand, experimental bounds on parity violation were used to calculate upper bounds on $\lambda_e, \lambda_q,$ and these bounds  were found to be ${\cal  O}(\sqrt{G_F})$~\cite{Chakraborty:2024zek}. Keeping these in mind, we have taken $\lambda^2= \pm 0.1~ G_F\,$ for the purpose of comparison with the standard interaction (SI, all $\lambda=0$).  As we mentioned earlier, we have written $\lambda^2$ to represent $\lambda \Delta \lambda$\,, or more precisely $\lambda_e\Delta\lambda_{ij}\,,$ so here $\lambda^2<0$ means $\Delta \lambda_{21}, \Delta\lambda_{31} <0\,.$

\begin{figure}[thbp]
\includegraphics[width=7.5cm]{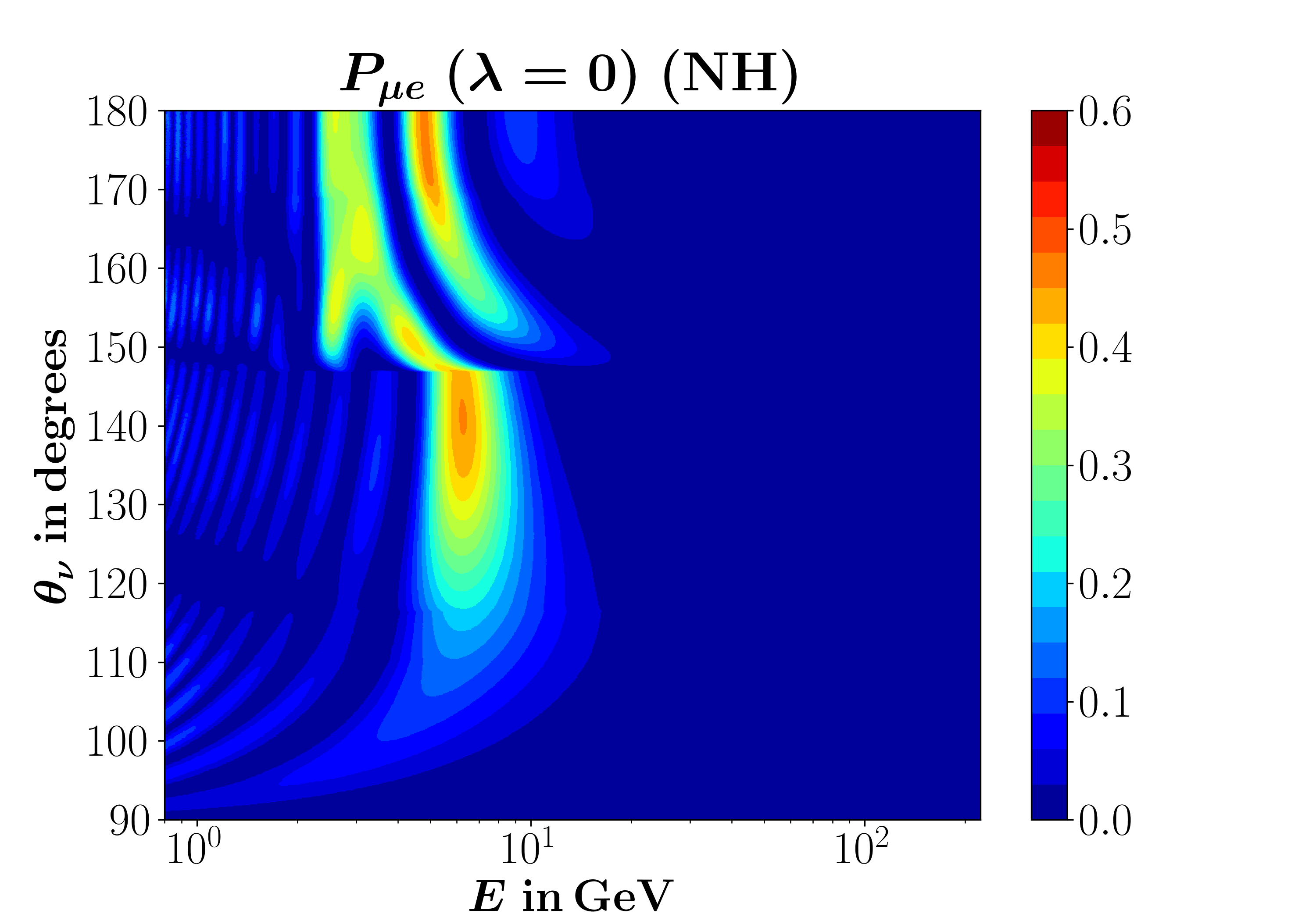}
 \hspace{0.5cm}
 \includegraphics[width=7.5cm]{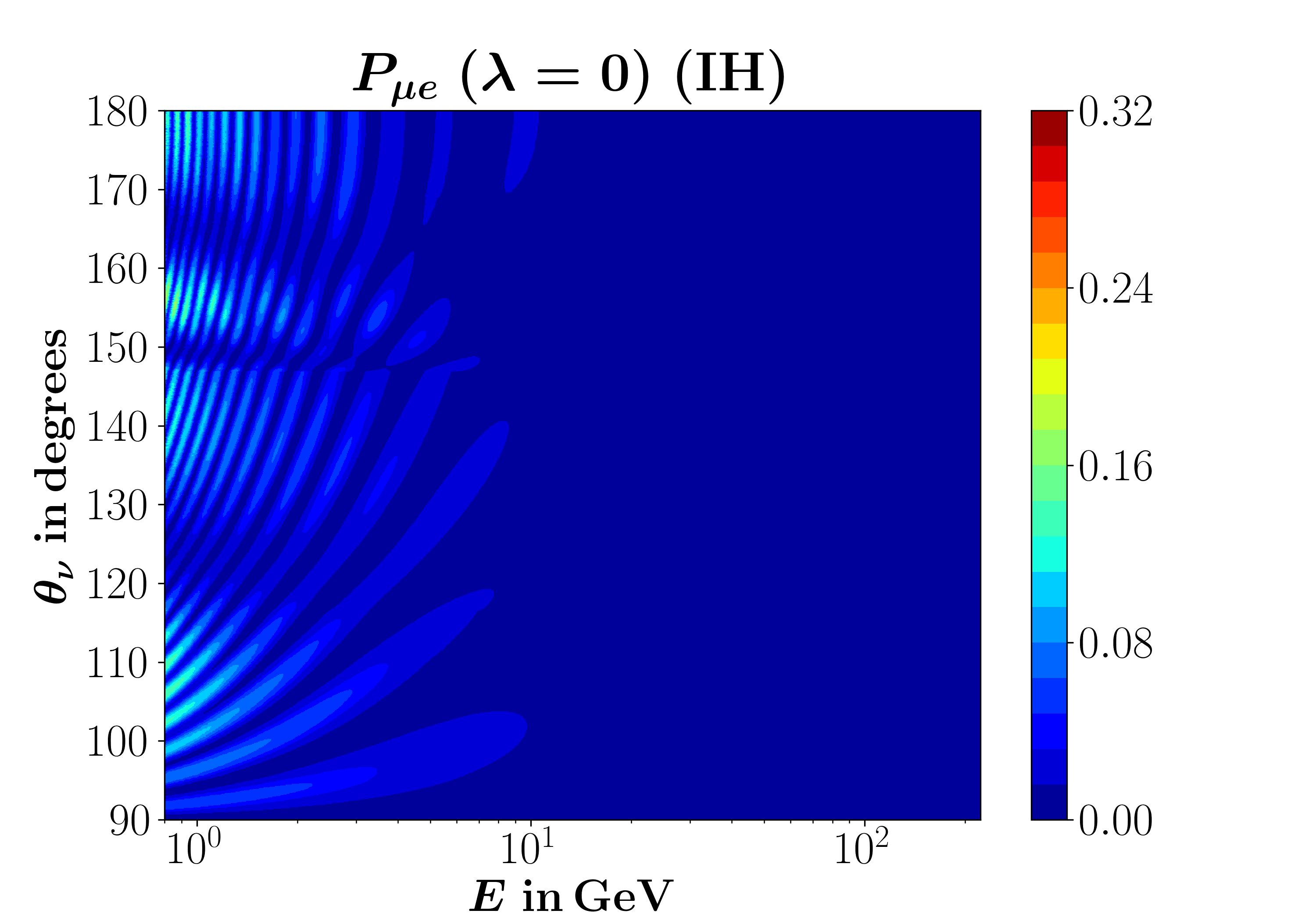}
 \includegraphics[width=7.5cm]{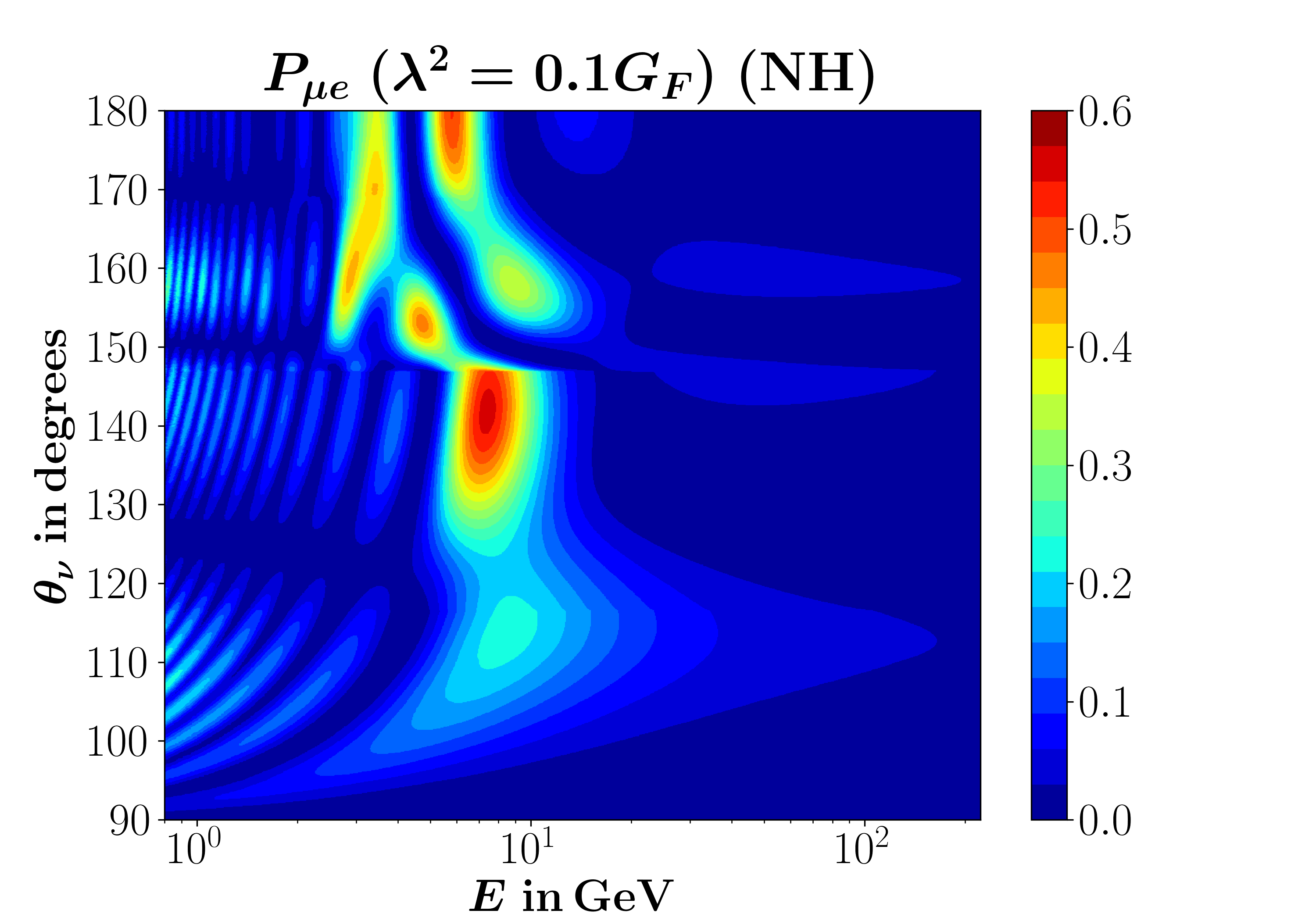}
 \hspace{0.5cm}
 \includegraphics[width=7.5cm]{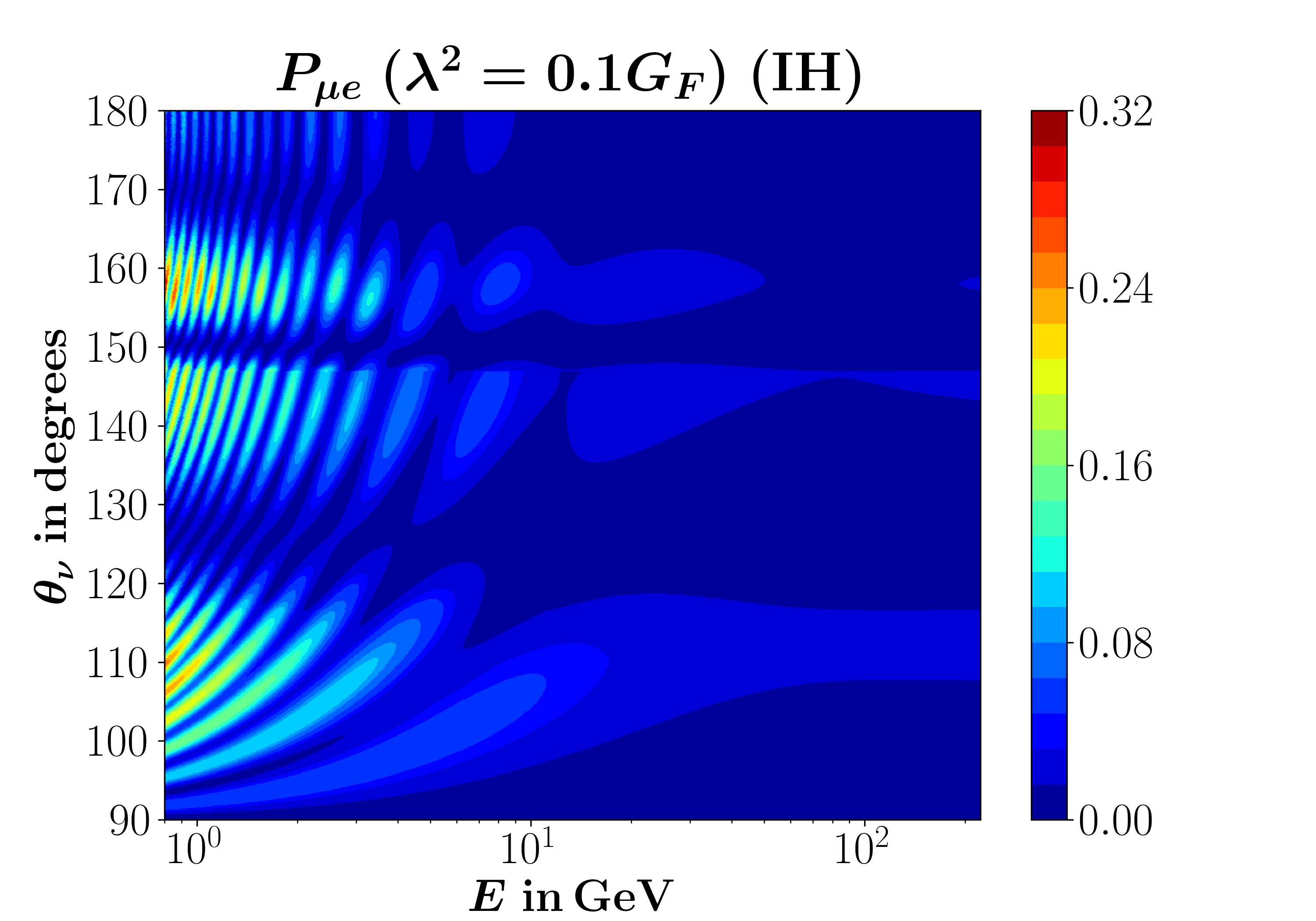}
 \includegraphics[width=7.5cm]{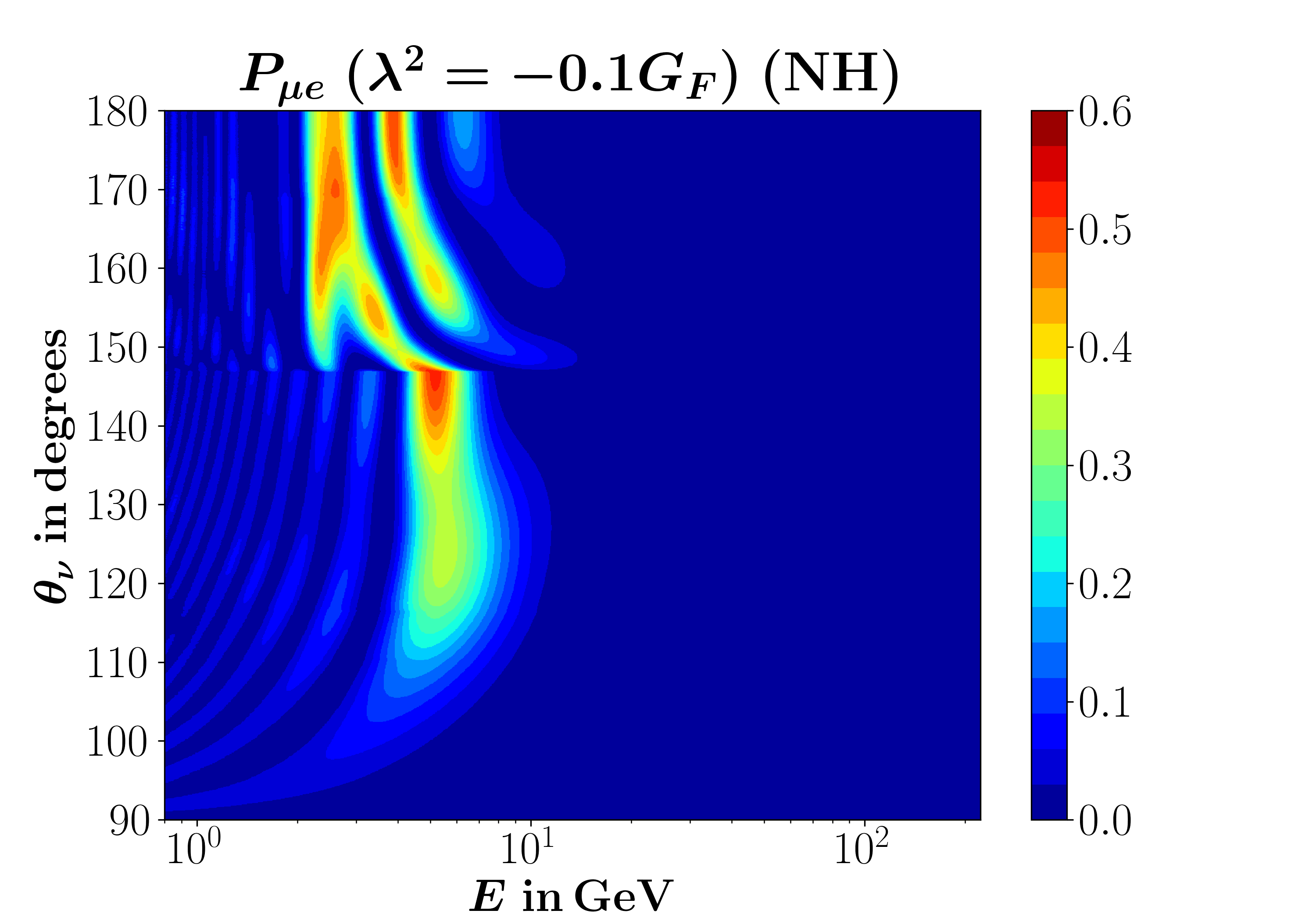}
\caption{Oscillograms for the $\nu_\mu \to \nu_e$ conversion probability for NH and IH. 
The values of $\lambda^2$ used for these are mentioned at the top of the oscillograms.
} 
\label{oscillogram-e}
\end{figure}
%

\subsection{{$P(\nu_\mu \to \nu_e)$}}
Let us now discuss the oscillograms. We will first focus on those for $P_{\mu e}$. The top left panel of Fig.~\ref{oscillogram-e} shows the dependence of $P_{\mu e}$ on the energy $E\,$ and the zenith angle $\theta_\nu\,$ for Standard Interaction (SI) with normal hierarchy (NH). {We can see that} in this case $P_{\mu e}$ experiences both the MSW resonance as well as parametric resonance.
The MSW resonance pattern is observed for neutrinos which cross only the mantle, with the main peak around $125^\circ<\theta_\nu<147^\circ$ and 5.5 {GeV}$ <E< 8$ GeV\,, and is denoted by the red and yellow patches within the green band. On the other hand, {for the core crossing neutrinos}, the MSW resonance within the core is also observed at $\theta_\nu>147^\circ$ and 2~GeV$ < E < 3~\rm GeV\,,$  while the parametric resonance can be identified as the yellow and red patches at the zenith angle $\theta_\nu > 147^\circ$ and energy $3\,\mathrm{GeV}<E<6.5\,$ GeV,   {as expected}~\cite{Qian:2015waa,Chatterjee:2014gxa,Esmaili:2013fva,Kumar:2021faw,Bakhti:2022axo,Akhmedov:2006hb, ESSnuSB:2024wet,Sahoo:2023mpj,Upadhyay:2024gra,Bernabeu:2001xn,Akhmedov:2008qt}. MSW resonance takes place both in the mantle and core region, whereas parametric resonance occurs because of the periodic nature of Earth's density profile --- in particular the mantle-core-mantle transition 
{(see the discussion in Sec.~\ref{resonance-derivation})}. We get two sharp transitions {or discontinuity lines at the zenith angle} $\theta_\nu \sim 117^\circ$ and $147^\circ\,$ --- the first
one at $\theta_\nu \sim 117 ^\circ$ occurs at the boundary of crust and mantle and the other one at $\theta_\nu \sim 147^\circ$ is due to the presence of core-mantle boundary. Thus one can get an idea about the Earth's tomography from the atmospheric neutrino oscillation study.

Let us next consider the middle left panel of Fig.~\ref{oscillogram-e}, {which shows the $P_{\mu e}$ oscillogram for} $\lambda^2 = 0.1~ G_F$ 
and for normal mass hierarchy.
The MSW resonance {peak within the mantle} can be seen around 6~GeV$<E< 10$~GeV and $125^\circ <\theta_\nu< 147^\circ$\,, while the parametric resonance {ridges} {appear} within the energy range of $3~\mathrm{GeV} <E< 11~\mathrm{GeV}$ and zenith angle $\theta_\nu > 147^\circ$. {The MSW peak within the core bends slightly towards the lower energy region.} The difference with the SI case is that the $\nu_\mu \to \nu_e$ conversion probability increases for $\lambda^2 > 0\,,$ {more so at higher energies}. In the MSW resonance region within the mantle, the red patch increases, and in the core region red patches also appear {in the MSW and parametric resonance ridges.} 
{Another obvious difference is that here we get non-zero $P_{\mu e}$ in the region $20~\mathrm{GeV}<E<200~\mathrm{GeV}$ but for the SI case $P_{\mu e}\approx0$ in this energy range.}

We now come to the $P_{\mu e}$ oscillogram for $\lambda^2=-0.1~G_F$ and with NH, which is shown in the bottom panel of Fig.~\ref{oscillogram-e}. Here also the conversion probability is different compared to the SI case. The red and yellow patches in the MSW resonance region {within the mantle are smaller now}. On the other hand, {in the parametric and MSW resonance regions} in the core, the red patches within the yellow band {are bigger now}, especially in the band near $E\sim 3~\mathrm{GeV}\,.$ Here also the oscillogram pattern changes but it is different from the $\lambda^2 = 0.1~ G_F$ case. Here the whole bands of the oscillogram shrink slightly compared to the SI case. Similar to the SI, here also we find $P_{\mu e} \approx 0$ in the region $E>20$~GeV. {We can understand this by noting that $2\tilde{n}\Delta\lambda E \equiv 2n\lambda^2 E$ adds to $\Delta m^2$ in the oscillation formulae, so the effect of $\lambda^2$ is reduced for large $E$ when $\lambda^2<0\,.$  }

For inverted hierarchy (IH), the top right and middle right panels in Fig.~\ref{oscillogram-e}  do not show the MSW or parametric resonance. However the $\nu_\mu \to \nu_e$ conversion probability increases for $\lambda^2=0.1~ G_F$ compared to the SI case. The discontinuity line at $\theta_{\nu}\sim 147^\circ$ remains visible for both the cases.

\begin{figure}[thbp]
\includegraphics[width=7.5cm]{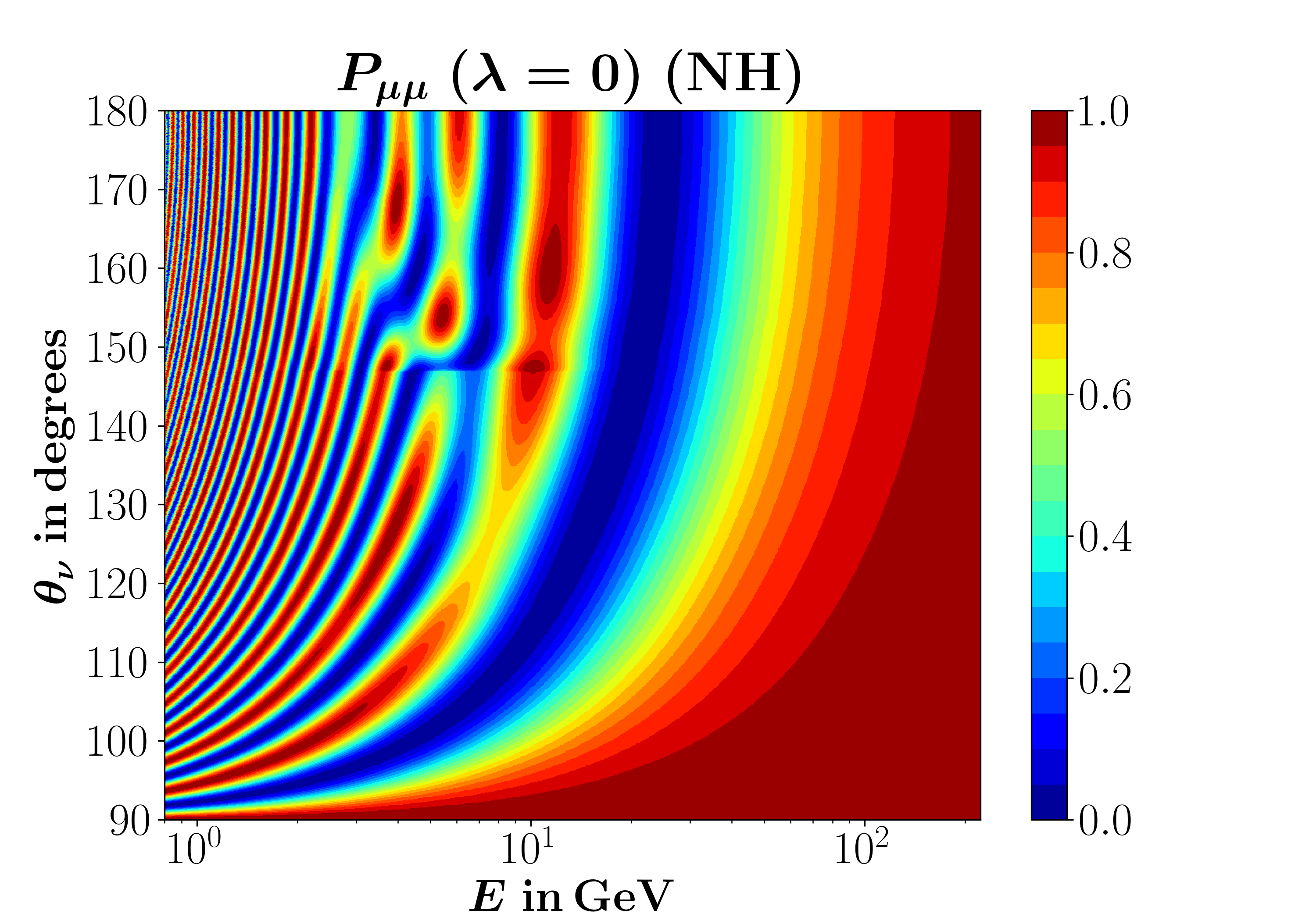}
\hspace{0.5cm}
\includegraphics[width=7.5cm]{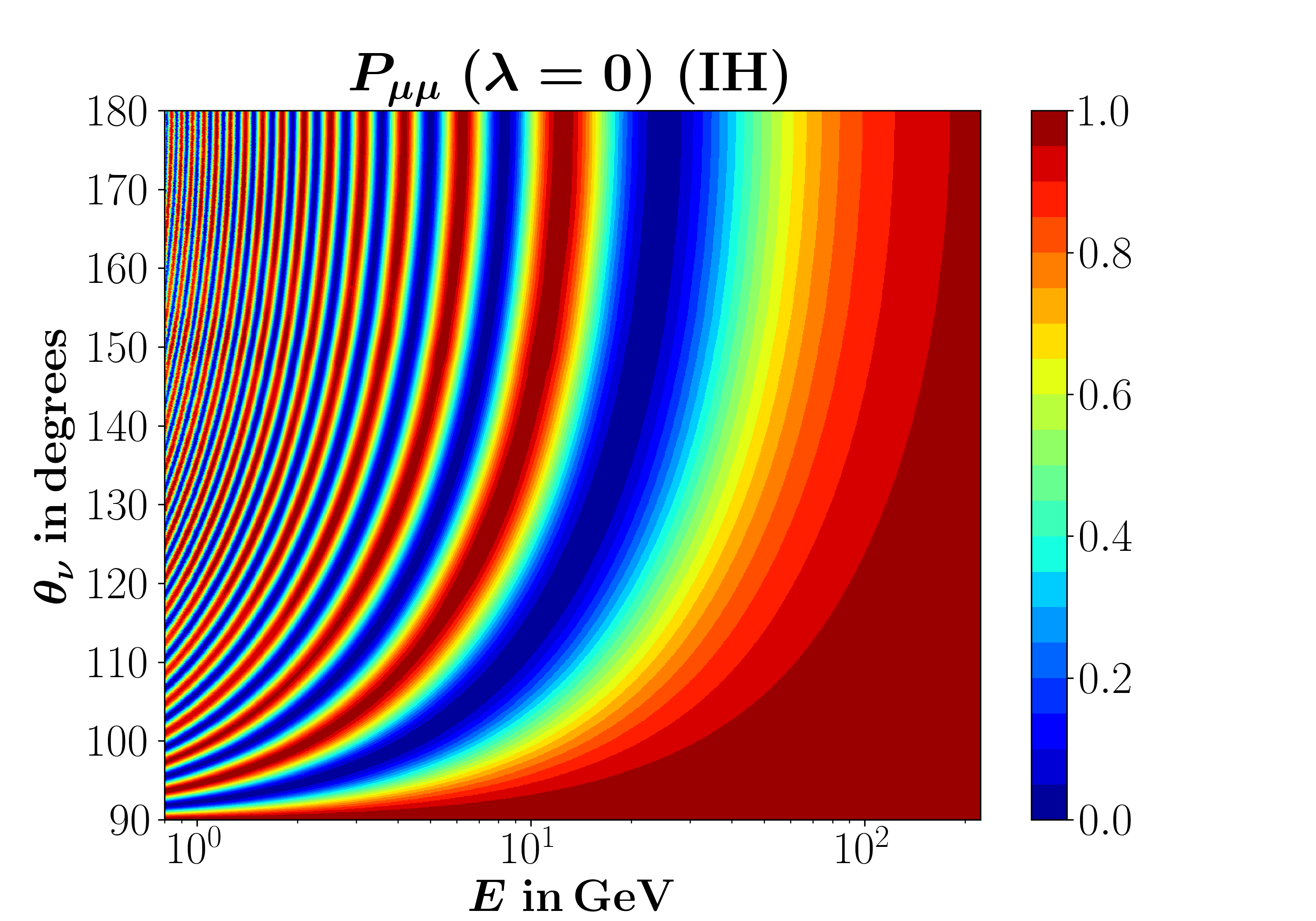}
\includegraphics[width=7.5cm]{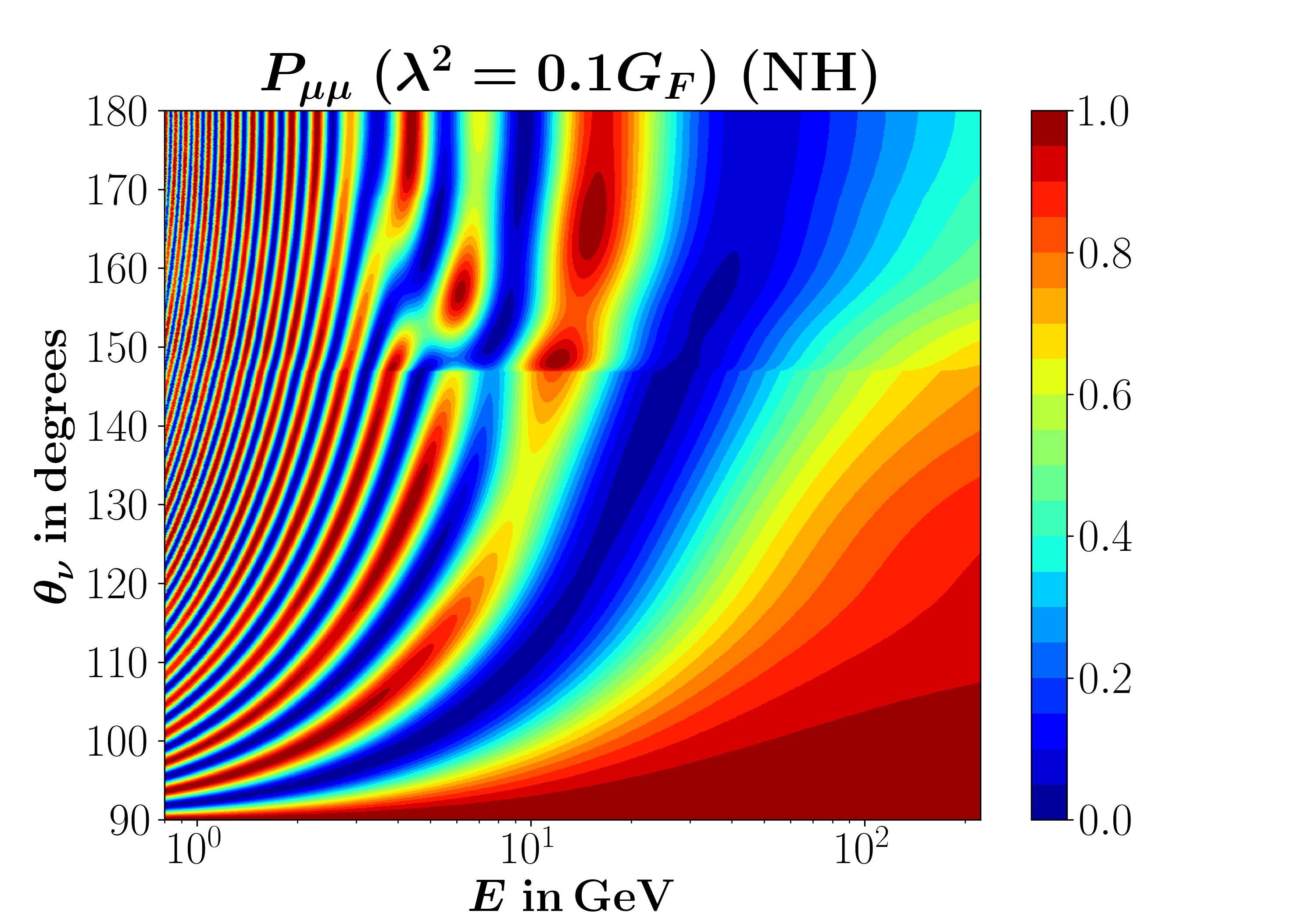}
\hspace{0.5cm}
\includegraphics[width=7.5cm]{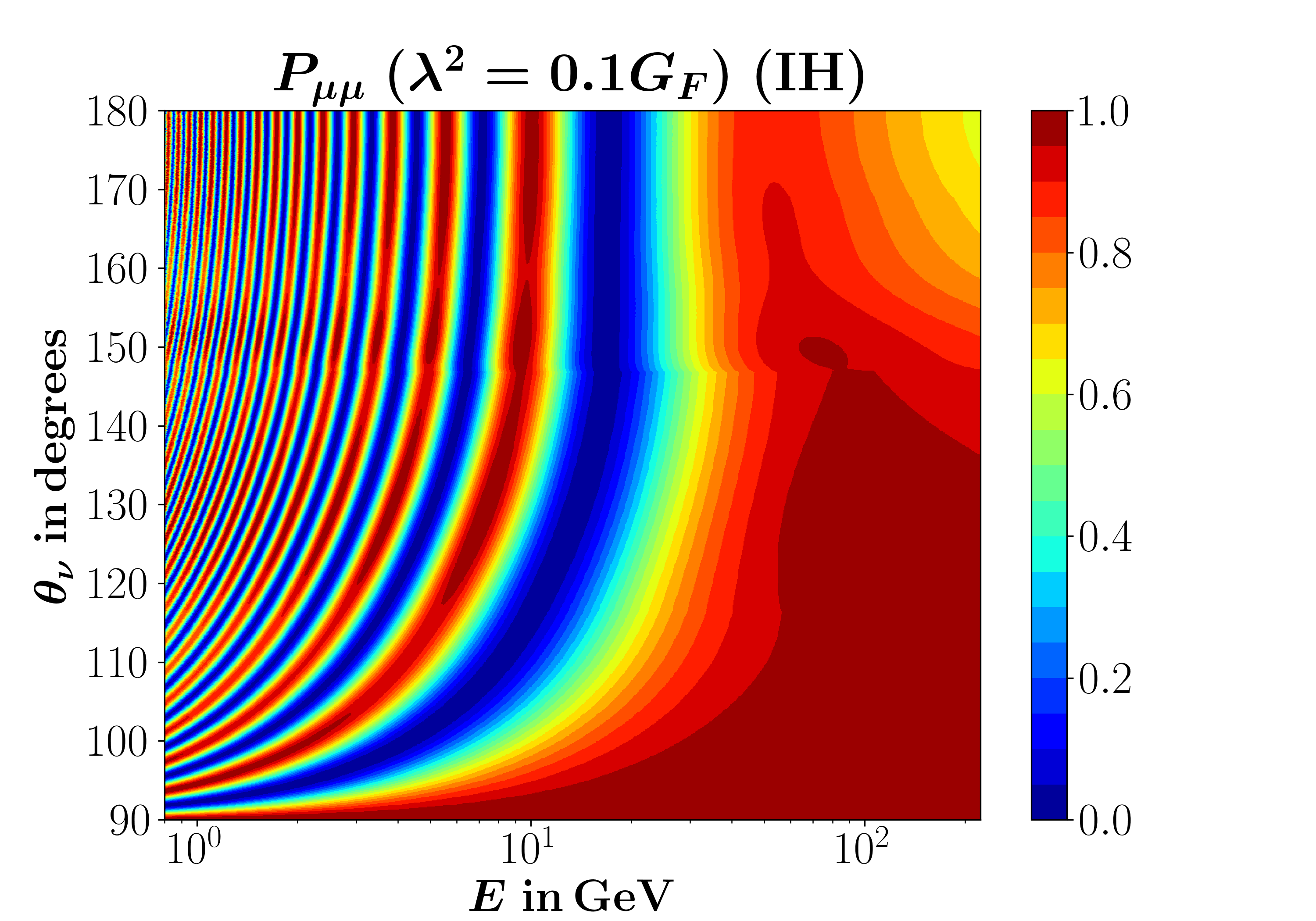}
\includegraphics[width=7.5cm]{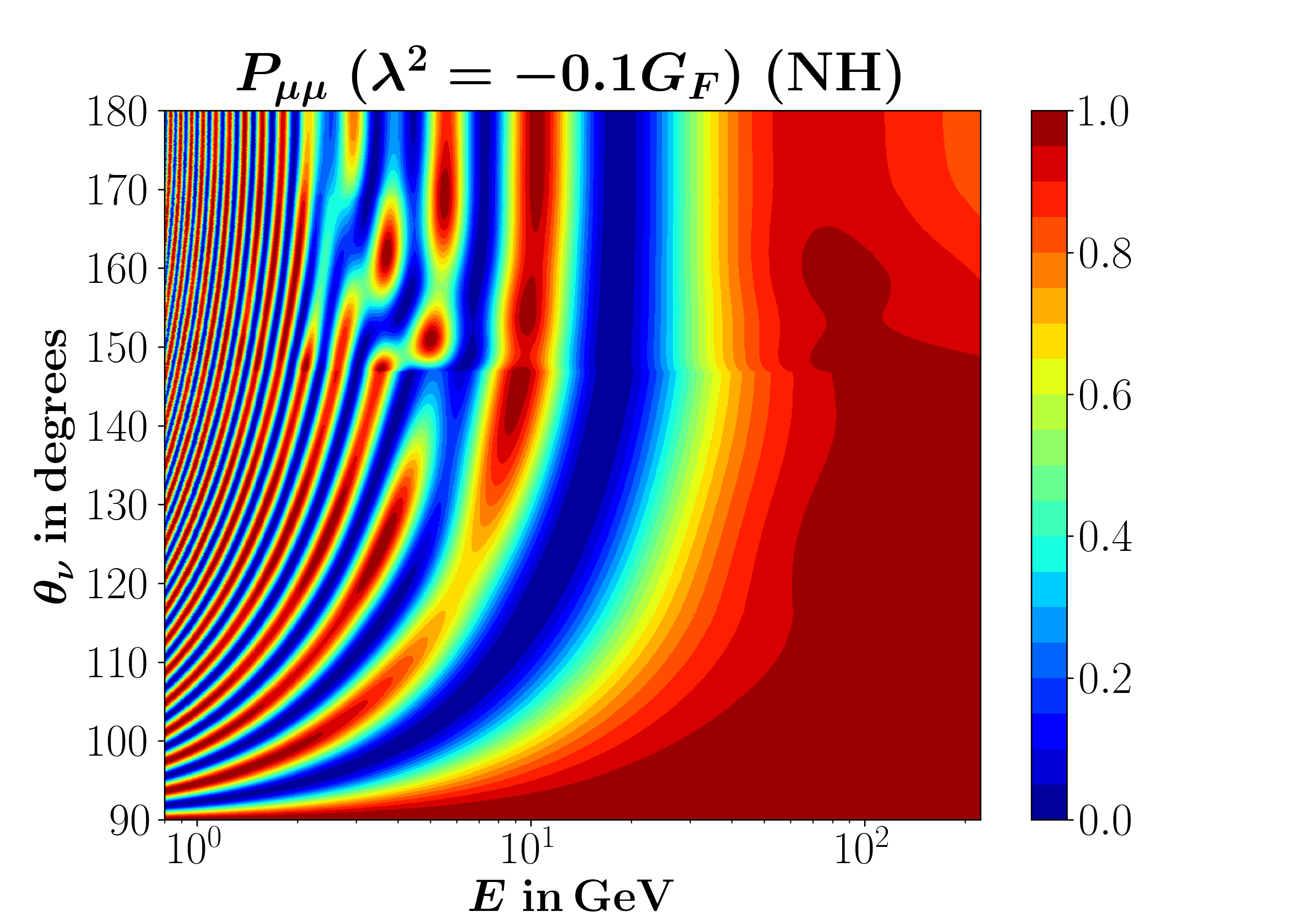}
\caption{Oscillograms for $\nu_\mu$ survival probability for SI and including the effect of the geometrical quartic interaction {for both NH and IH}. 
}
\label{oscillogram-mu}
\end{figure}

\subsection{{$P(\nu_\mu \to \nu_\mu)$}}

Let us now consider the oscillograms for $\nu_\mu$ survival probability, shown in Fig.~\ref{oscillogram-mu}.
The top two panels of Fig.~\ref{oscillogram-mu} represent the SI case, the middle two panels correspond to $\lambda^2 = 0.1~ G_F$, and the bottom panel corresponds to $\lambda^2 = -0.1~ G_F$. The dark blue band, which starts near $\theta_\nu = 90^\circ$ and ends at $\theta_\nu = 180^\circ$, is where the $\nu_\mu$ survival probability is at its lowest and is called an oscillation valley \cite{Kumar:2021lrn,Kumar:2020wgz}. This valley exists in the zone {$E < 30\mathrm{\, \rm GeV}$} for SI. {The higher order oscillation minima are represented by the thinner blue bands which exist up to $ 45$~GeV} beyond which the survival probability increases {rapidly} {in the higher energy region}. In contrast, for $\lambda^2=0.1~ G_F$ and NH, {the oscillation valley (the dark blue band) shrinks and extends to $E < 45~\rm GeV$ but only for $90^\circ<\theta_\nu<162^\circ\,.$ While the higher order minima bands are quite a bit wider and extends up to 200 GeV}. 
But for $\lambda^2=-0.1~ G_F\,$ and NH, the oscillation valley is present only for {$0.7~\mathrm{GeV} < E < 25~\mathrm{GeV}$}. In the top left panel, the region between two dark red patches, where a yellow band with a very light red patch is visible, between $6~\mathrm{GeV}<E<10~\mathrm{GeV}$ and {$118^\circ<\theta_\nu<132^\circ$} is the MSW resonance region. The parametric resonance for the SI case is present around {3.5~{GeV}$ < E < 7$ {GeV} and $\theta_\nu>147^\circ$} {as is seen} by three red patches. In the middle left panel {($\lambda^2 = 0.1 ~ G_F $ and NH)}, we can see that the MSW resonance region stretches and in the parametric resonance region two red patches are seen. In the bottom panel ($\lambda^2 = -0.1~ G_F$ and NH), the MSW resonance region has shrunk, while in the parametric resonance region, there are also three red patches similar to the SI case.  {For IH, no resonance pattern is present for $\lambda^2=0$ and $\lambda^2= 0.1~ G_F$. But in presence of geometrical interaction $(\lambda^2=0.1~ G_F)\,,$ $P_{\mu \mu}$ changes, especially in the high energy region within the core. These plots are shown in the top and middle right panel of Fig.~\ref{oscillogram-mu}.}    These figures can also be understood as before by noting that $2n\lambda^2 E\,$ is added to $\Delta m^2\,$ in the oscillation formulae. 
\subsection{$P(\nu_\mu \to \nu_\tau)$}
\begin{figure}[hbtp]
\includegraphics[width=7.5cm]{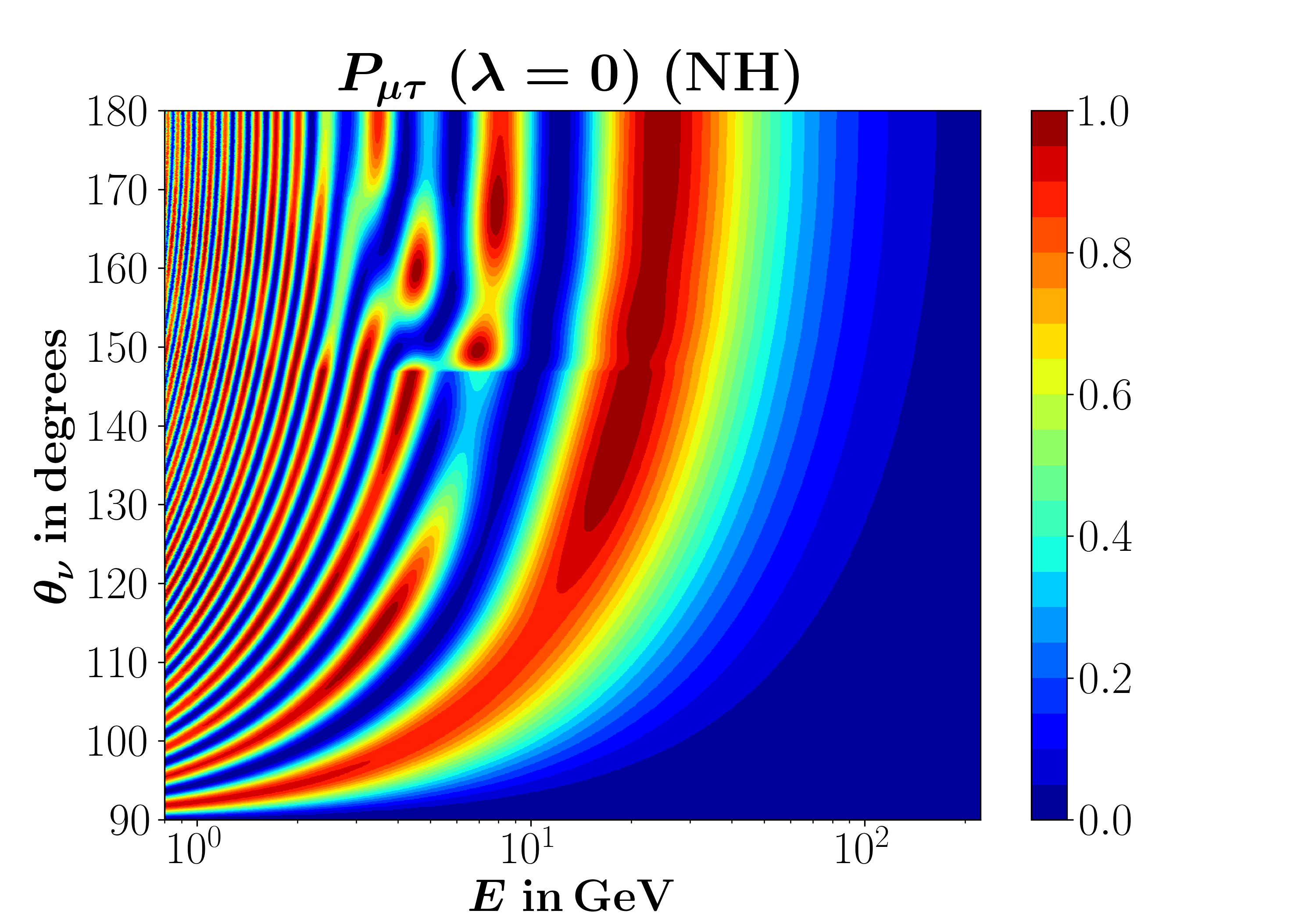}
\hspace{0.5cm}
\includegraphics[width=7.5cm]{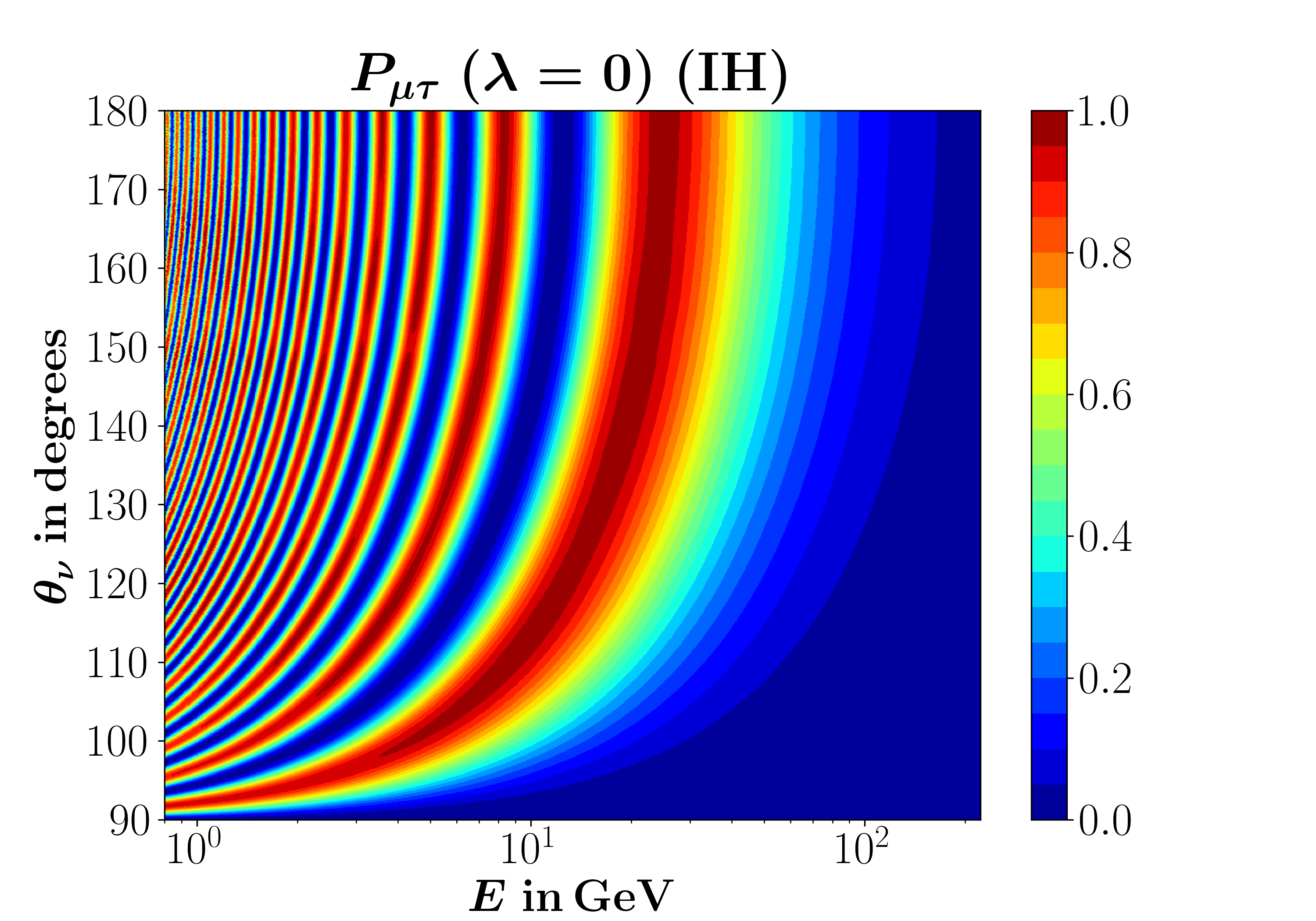}
\includegraphics[width=7.5cm]{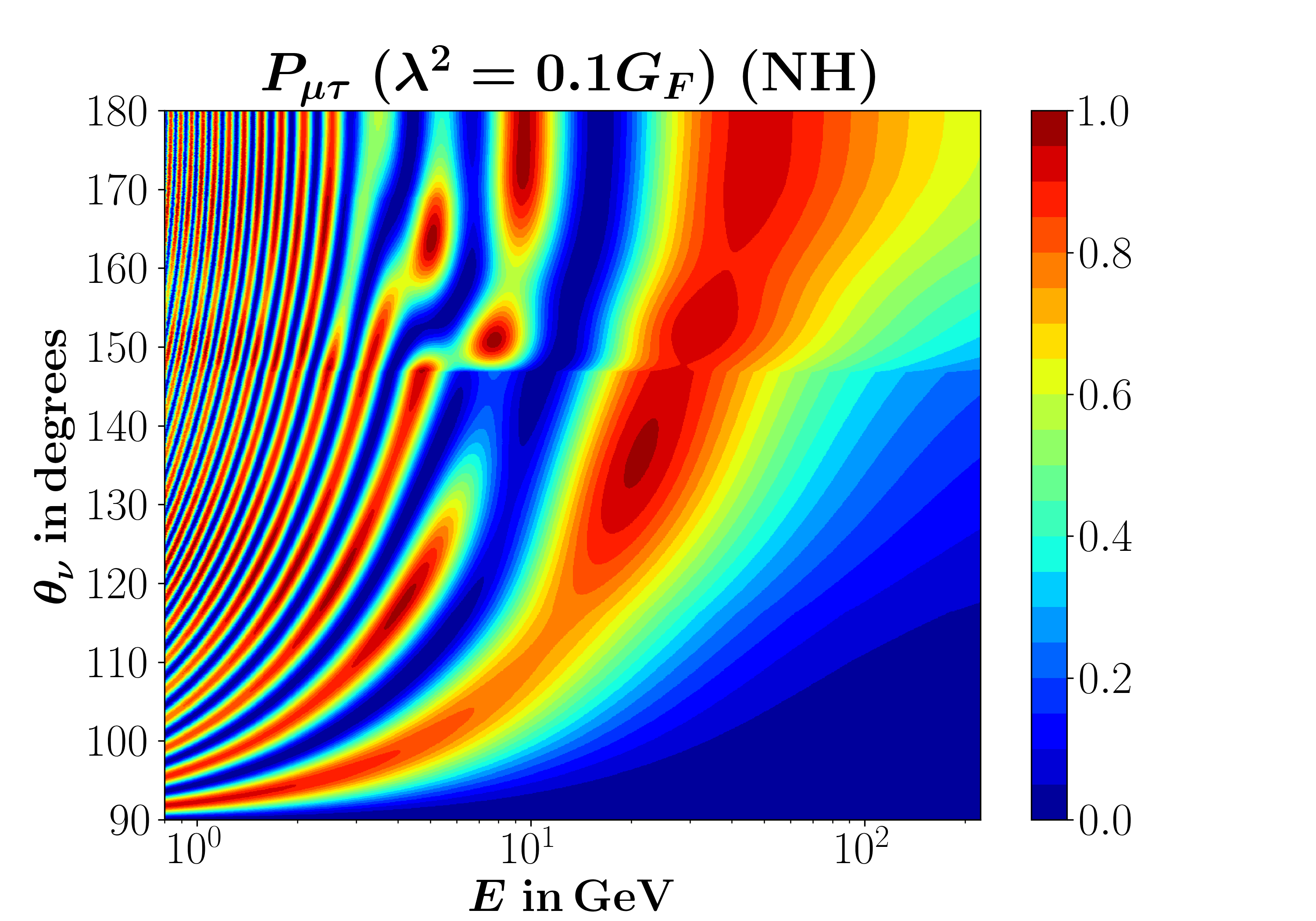}
\hspace{0.5cm}
\includegraphics[width=7.5cm]{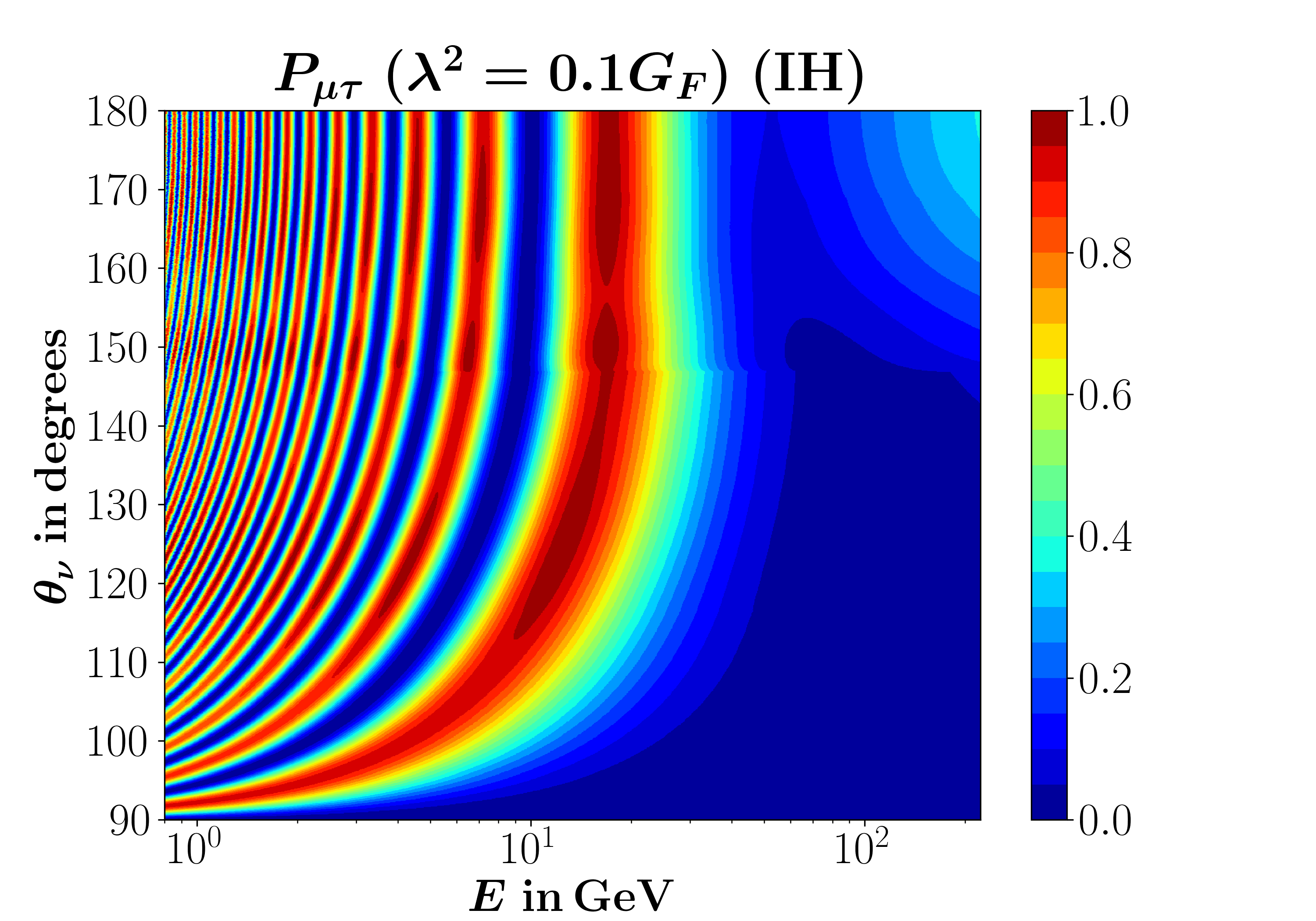}
\includegraphics[width=7.5cm]{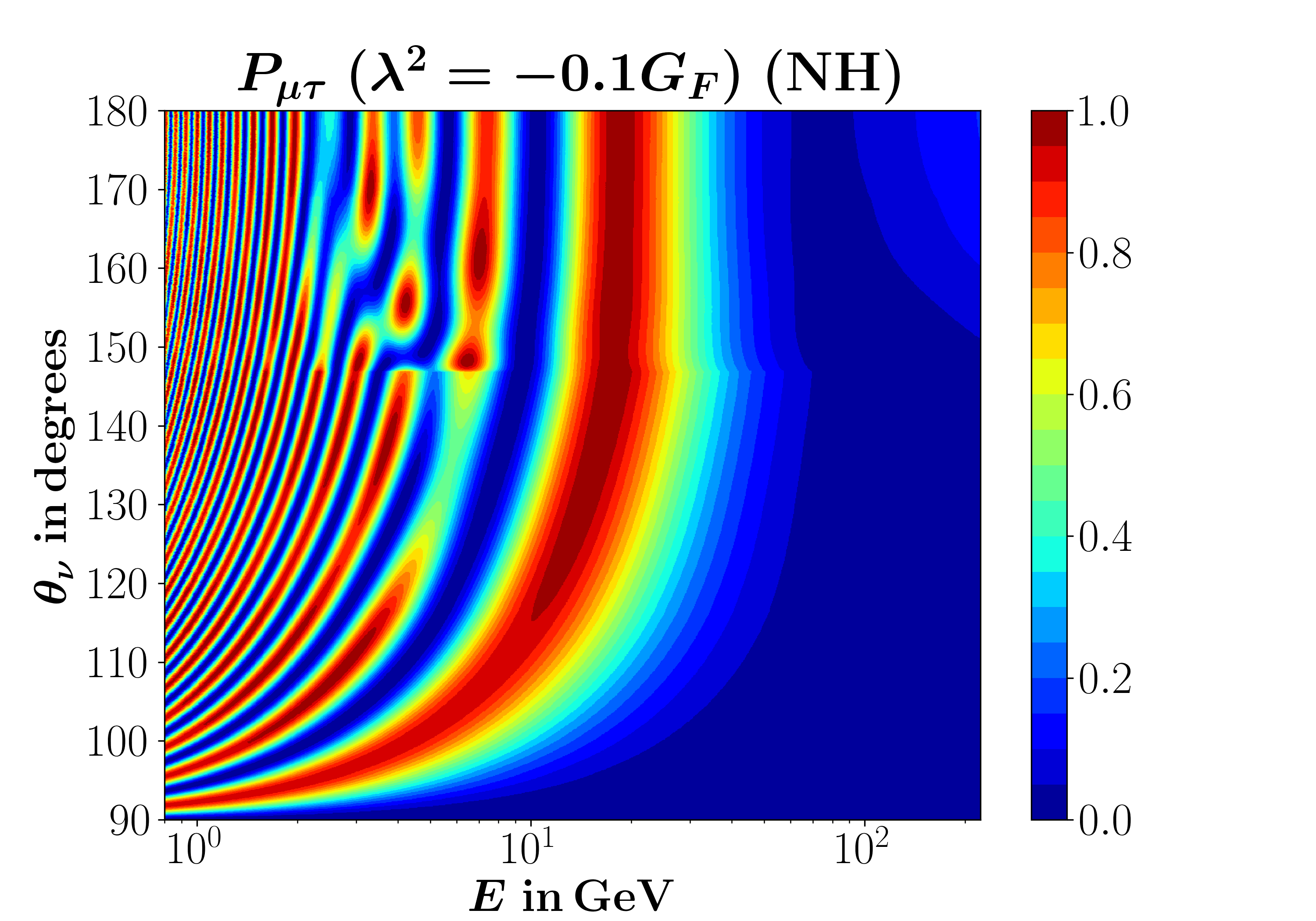}
\caption{Oscillograms for $\nu_\mu \to \nu_\tau$ conversion probability for SI and including the effect of torsional four-fermion interaction for both NH and IH.
}
\label{oscillogram-tau}
\end{figure}

{Let us also consider, for the sake of completeness,} the impact of this torsional interaction on $\nu_\mu \to \nu_\tau$ conversion. This is shown in Fig.~\ref{oscillogram-tau}. As in the other oscillograms, the red shaded band represents the region of maximum probability, in this case for conversion from $\nu_\mu$ to $\nu_\tau$. For the SI case {and NH,} this red band exists up to 40 GeV energy, whereas for $\lambda^2 = 0.1~ G_F$ {and NH,} it extends up to 180 GeV. But for $\lambda^2=- 0.1~G_F$ and NH, this red band exists up to $E\sim25$ GeV. This is also clearly related to the behavior of the $\nu_\mu$ survival rates.

{The anti-neutrino oscillation probabilities can be calculated from Eq.~(\ref{eq:TDSE_for_3nu.b}) by replacing $U \to U^*\,,A \to -A \,,$ {and} $\Delta \tilde{m}^2_{ij} := \Delta m^2_{ij} - 2 \tilde{n} E \Delta\lambda_{ij}\,.$ Similarly, oscillograms can be drawn for anti-neutrino oscillation probabilities. Here the MSW and parametric resonances will be observed for inverted hierarchy (IH). }

Thus we conclude that the torsional four-fermion interaction has a more significant impact on atmospheric neutrino oscillations in the high-energy region when $\lambda \Delta \lambda$ is positive. In the resonance region, the oscillation probability visibly changes for both positive and negative values of $\lambda^2$. Additionally, the presence of the geometrical interaction alters the oscillogram patterns in both the appearance and disappearance channels. {Here we have focused more on how the oscillation patterns change depending on the size (order of magnitude) of the $\lambda_i\,.$ Thus we have created the oscillograms by considering only one value of $\Delta \lambda_{ij}\,.$ We have included the possibility that $\lambda_f\Delta\lambda_{ij}<0\,,$ which covers both 
possibilities $\lambda_f <0$ and $\Delta\lambda_{ij}<0\,.$ If we take different values for $\Delta \lambda_{21}$ and $\Delta \lambda_{31}\,,$ we should find finer structure in the oscillograms both from the hierarchy of $\lambda_i$ and its combination with the mass hierarchy, as we have done in recent works on DUNE~\cite{Barick:2025ahc,inprep:2025}. Also from that we should be able to probe the hierarchy in $\lambda$ analogous to the mass hierarchy of neutrinos.} 

\subsection{Changes in probabilities}
\begin{figure}[hbtp]
\includegraphics[width=7.5cm]{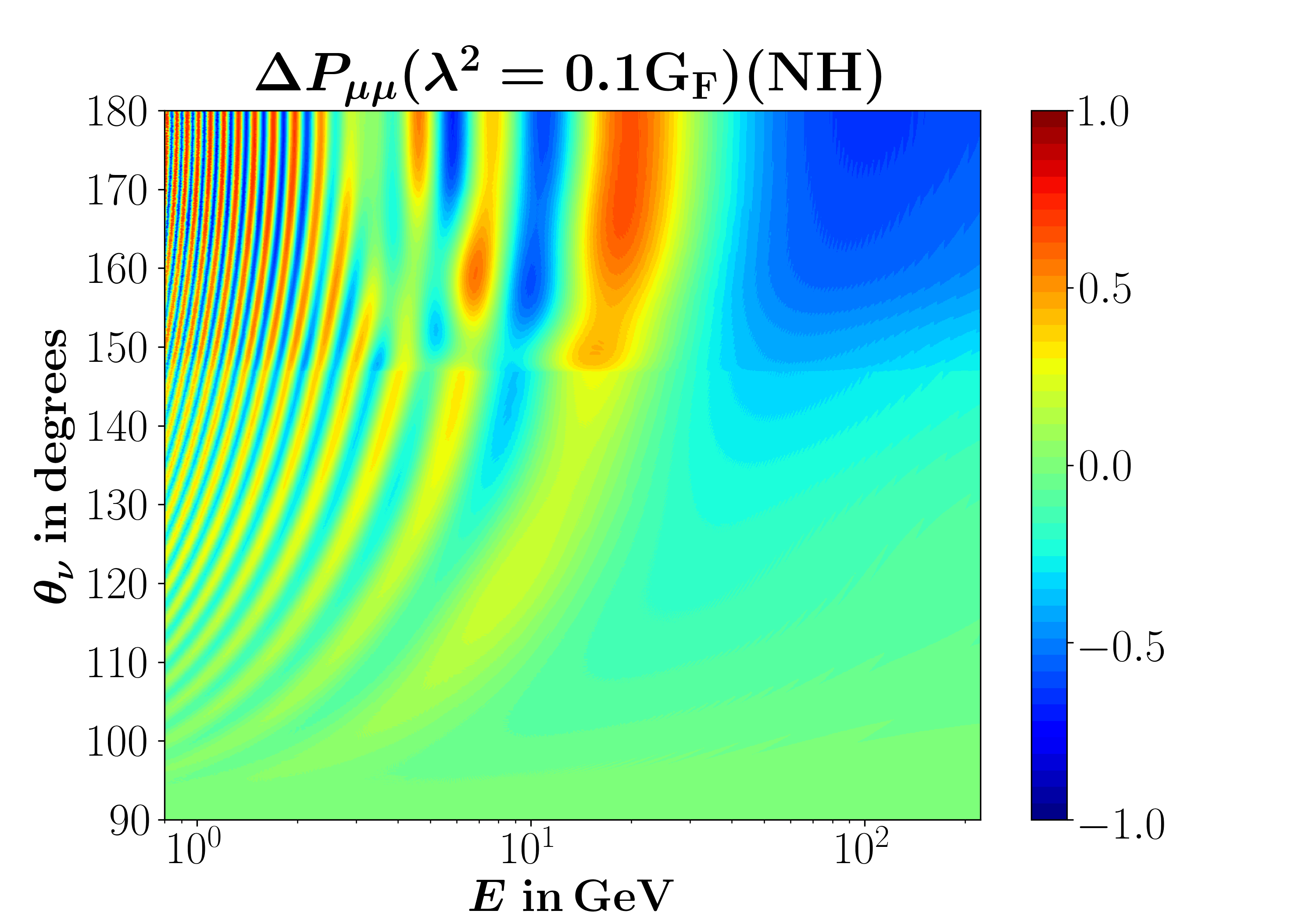}
\hspace{0.5cm}
\includegraphics[width=7.5cm]{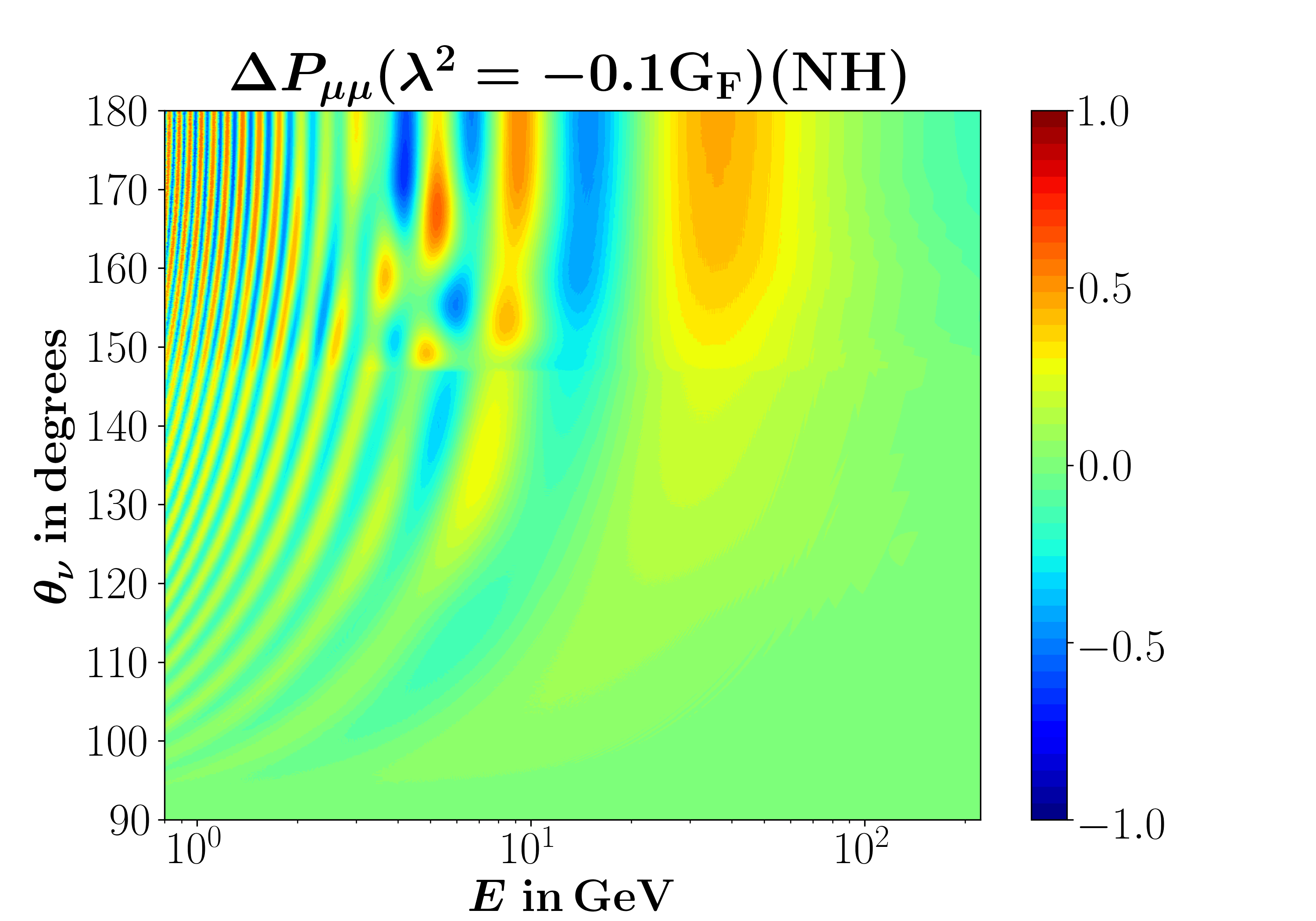}
\caption{The difference in the $\nu_\mu$ survival probability $\Delta P_{\mu \mu}=P^{\lambda\neq0}_{\mu \mu}-P^{\lambda=0}_{\mu \mu}$ in the $(E,\theta_\nu)$ plane for $\lambda^2=0.1~ G_F$ and $-0.1~ G_F\,.$}
\label{diff-mu-mu}
\end{figure}

\begin{figure}[hbtp]
\includegraphics[width=7.5cm]{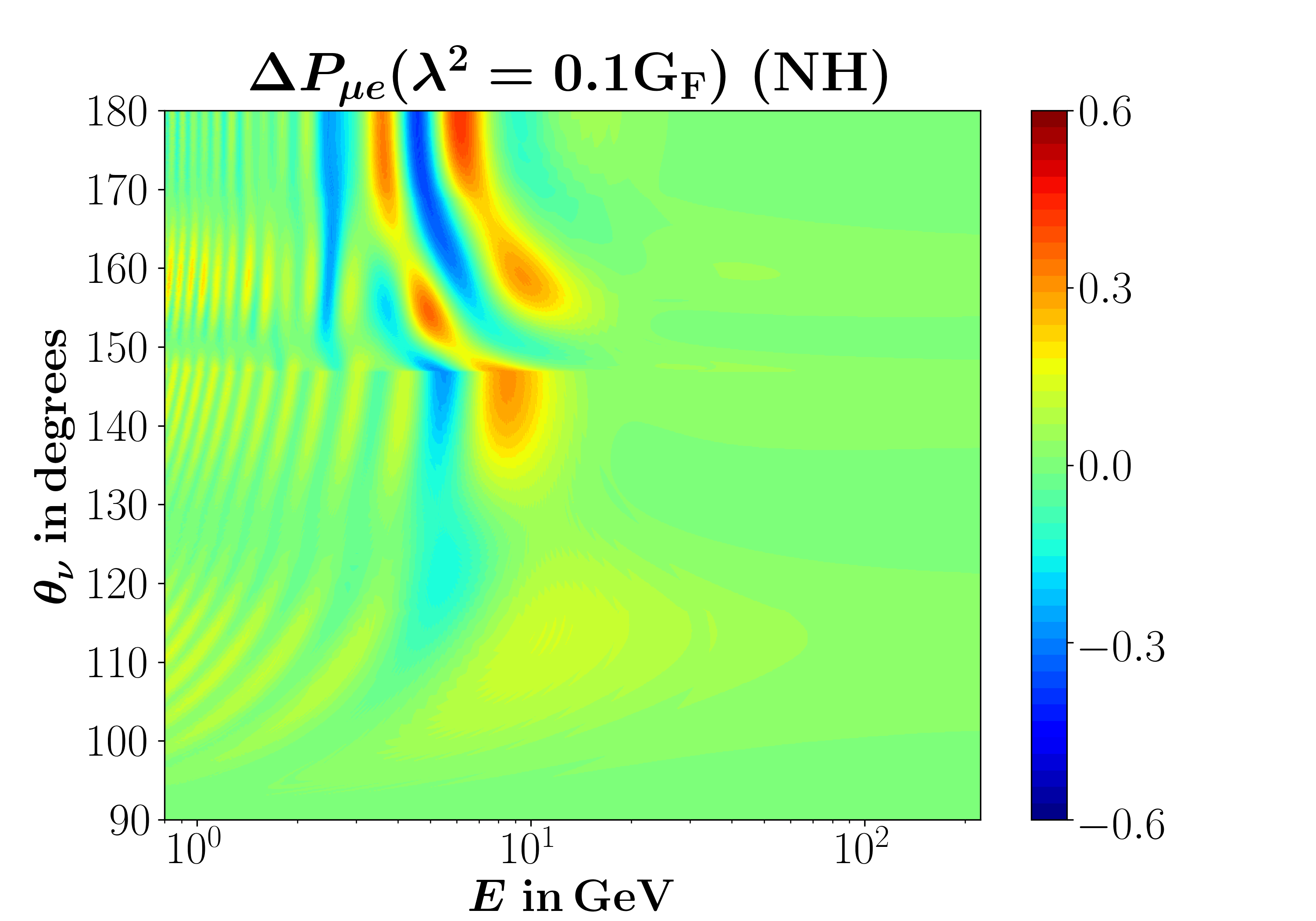}
\hspace{0.5cm}
\includegraphics[width=7.5cm]{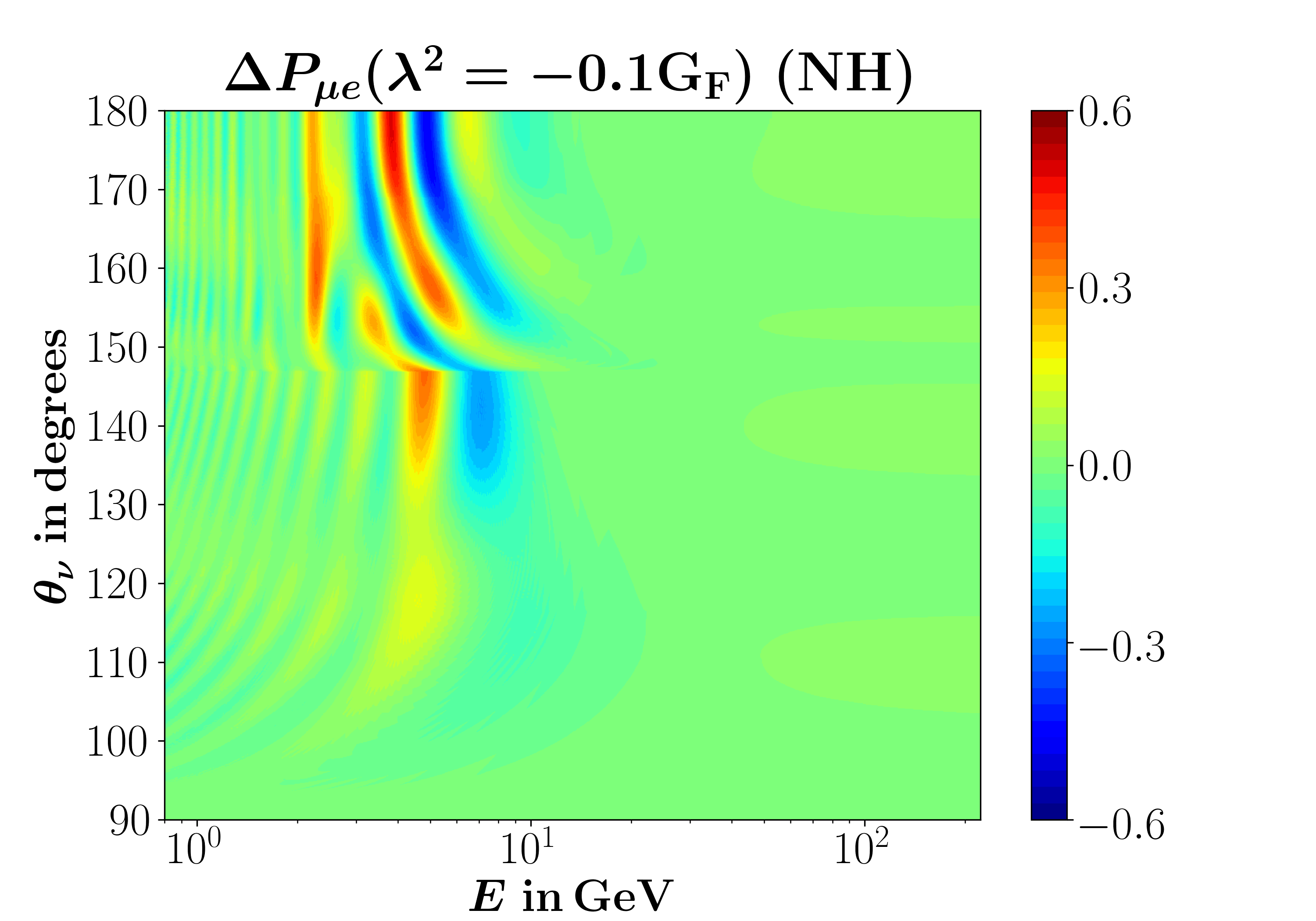}
\caption{The difference in the $\nu_\mu \to \nu_e$ conversion probability $\Delta P_{\mu e}=P^{\lambda\neq0}_{\mu e}-P^{\lambda=0}_{\mu e}$ in the $(E,\theta_\nu)$ plane for $\lambda^2=0.1~ G_F$ and $-0.1~ G_F\,.$}
\label{diff-mu-e}
\end{figure}

\begin{figure}[hbtp]
\includegraphics[width=7.5cm]{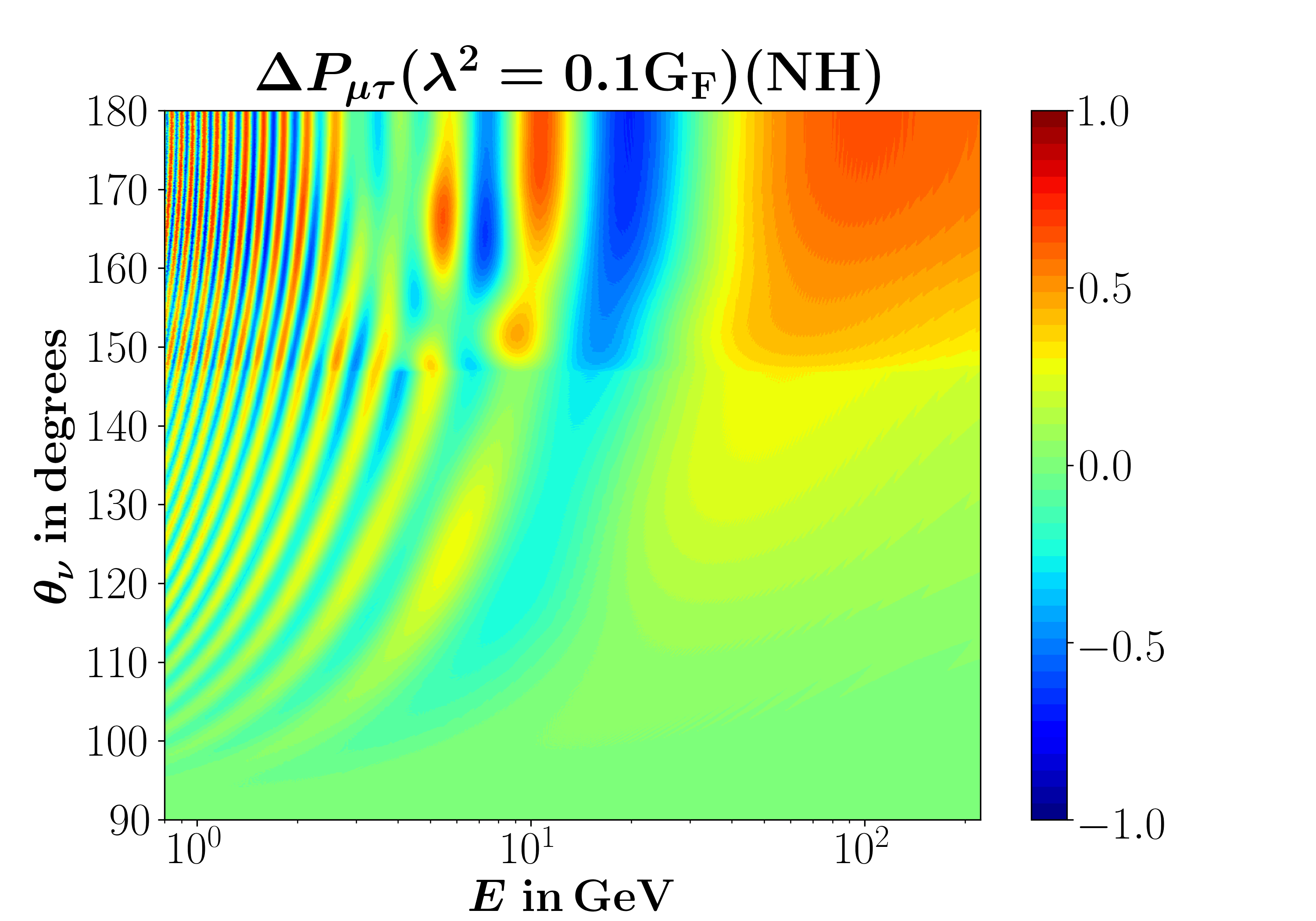}
\hspace{0.5cm}
\includegraphics[width=7.5cm]{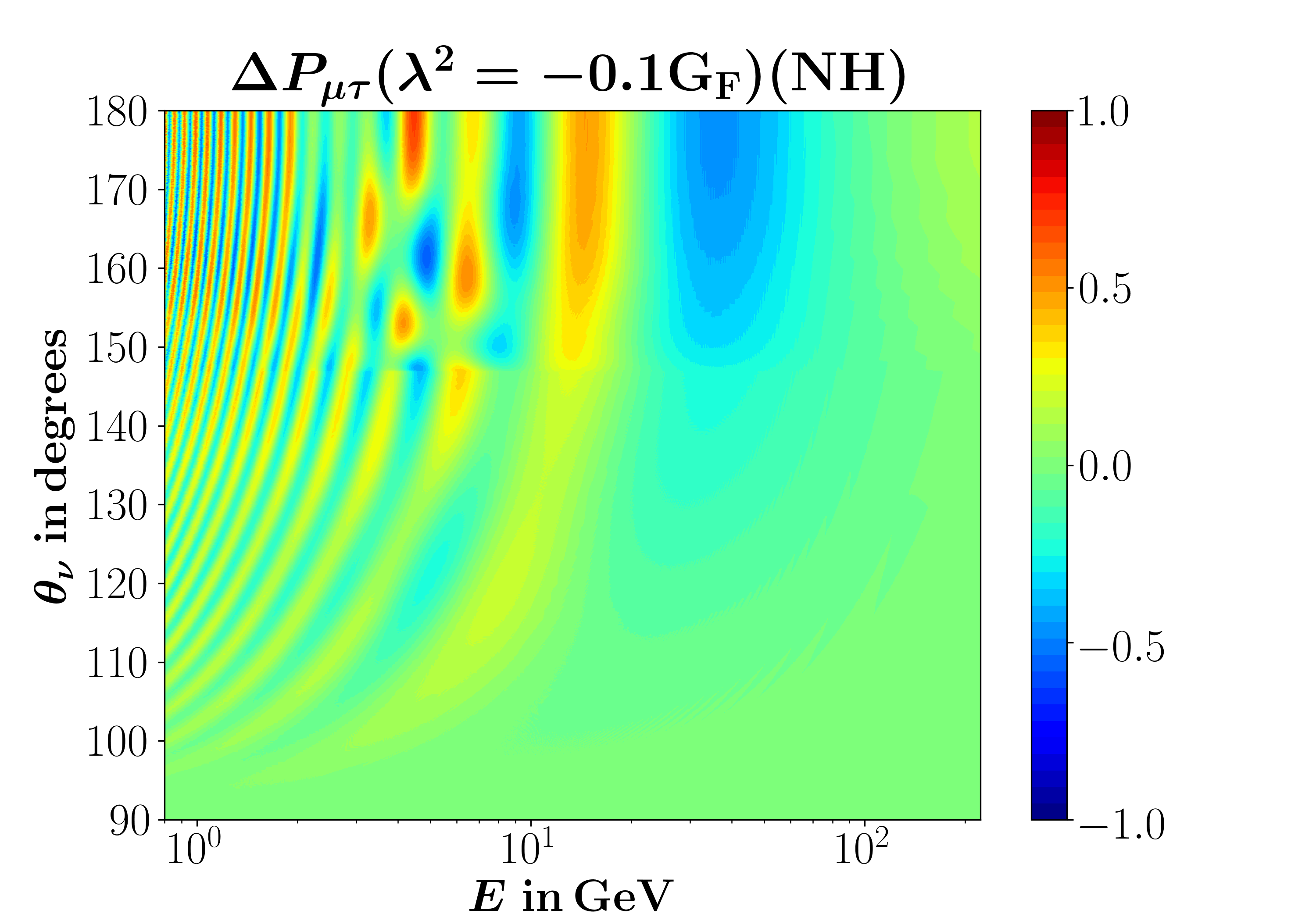}
\caption{The difference in the $\nu_\mu \to \nu_\tau$ conversion probability $\Delta P_{\mu \tau}=P^{\lambda\neq0}_{\mu \tau}-P^{\lambda=0}_{\mu \tau}$ in the $(E,\theta_\nu)$ plane for $\lambda^2=0.1~ G_F\,$and $-0.1~ G_F\,.$}
\label{diff-mu-tau}
\end{figure}

In Fig.~\ref{diff-mu-mu}, Fig.~\ref{diff-mu-e}, and Fig.~\ref{diff-mu-tau}, we show the change in survival and conversion probabilities, defined as $\Delta P = P^{\lambda \neq 0} - P^{\lambda = 0}$, in the $(E, \theta_\nu)$ plane. This helps in  understanding the impact of the geometrical four-fermion interaction on neutrino oscillation probabilities. Here $P^{\lambda = 0}$ represents the standard oscillation probability, while $P^{\lambda \neq 0}$ corresponds to the probability in the presence of the torsional interaction. It is evident from the figures that the most significant differences occur in the Earth's core region, with relatively smaller changes in the mantle. In particular, the left panels of Fig.~\ref{diff-mu-mu} and Fig.~\ref{diff-mu-tau} show that the probability is notably altered at higher energies within the core. {Qualitatively speaking, this is exactly as expected, because of the addition of $2\tilde{n}\Delta\lambda E$ to $\Delta m^2$ in the oscillation formulae.}

\begin{figure}[htbp]
\includegraphics[width=7.5cm]{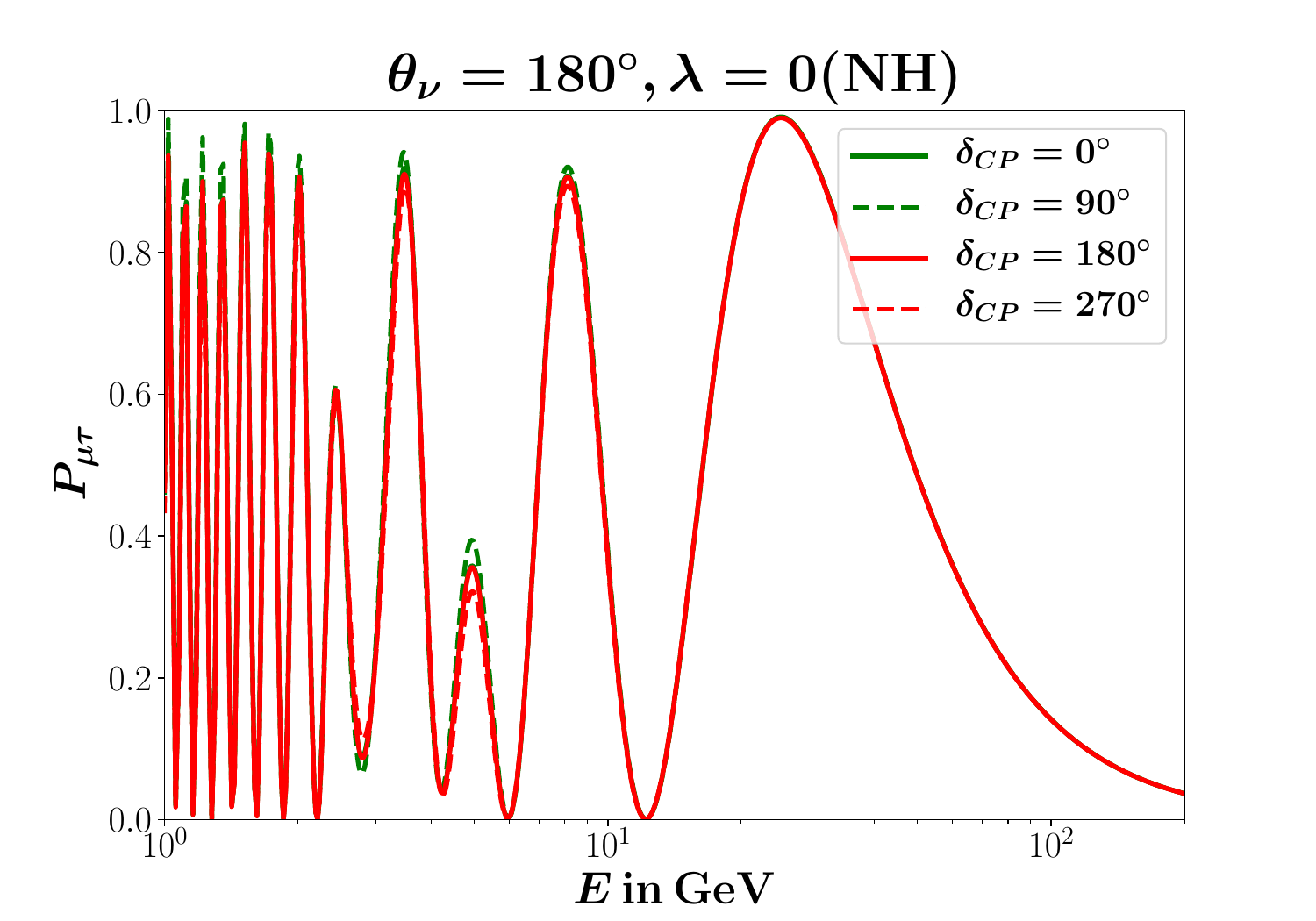}
\hspace{0.5cm}
\includegraphics[width=7.5cm]{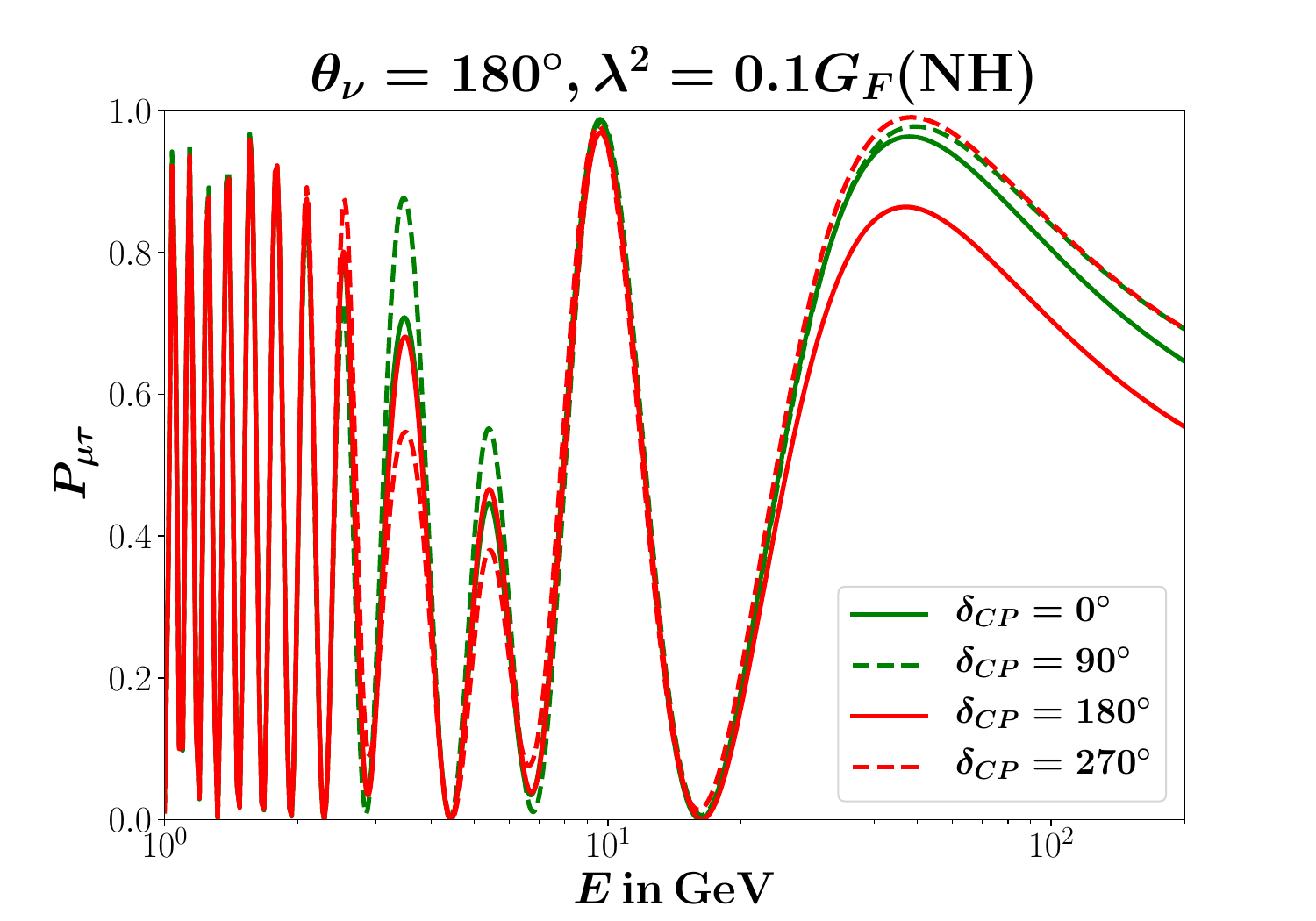}
\caption{$P_{\mu \tau}$ vs E for different values of $\delta_{CP} $ for a fixed zenith angle : $\theta_\nu=180^\circ$, for two values of the geometrical coupling constant $\lambda=0$ and $\lambda^2=0.1~ G_F$.}
\label{CP}
\end{figure}

{The quartic interaction violates parity as should be obvious from Eq.~(\ref{quartic-int}), so we should check if the CP-phase angle $\delta_{CP}$ affects the conversion and survival rates differently in its presence, compared to SI. }
Fig.~\ref{CP} shows the $\nu_\mu \to \nu_\tau$ conversion probability $P_{\mu \tau}\,$ as a function of the energy $E\,$ for four different values of the CP-phase angle  $\delta_{CP}$ and at a fixed zenith angle $\theta_\nu = 180^\circ$. The left panel is for SI  ($\lambda=0$) and the right panel is for $\lambda^2 = 0.1~ G_F$. 
{The other neutrino oscillation parameters used here are the same as those for the probability oscillograms in Fig.~\ref{oscillogram-e}--Fig.~\ref{diff-mu-tau}, for all of which the value of $\delta_{CP}$ was 232$^\circ$.} From the plot for the SI case, it is clear that $P_{\mu \tau}$ depends on $\delta_{CP}$ very loosely in the low energy region, while above $15$ GeV, $P_{\mu \tau}$ is completely independent of $\delta_{CP}$. On the other hand, when we include the effect of geometrical coupling with $\lambda^2 = 0.1~ G_F$\,,  we see that $P_{\mu \tau}$ depends on $\delta_{CP}$ even in the very high energy region $\sim$200 GeV. We {do not consider energies beyond 200 GeV} since the flux of the atmospheric neutrino reduces as $E^{-3.7}$~\cite{Gaisser:2002jj}.

%

\section{MSW and Parametric resonances}\label{resonance-derivation}
{In this section we will discuss how the MSW and parametric resonances in neutrino oscillations
are affected by the geometrical interaction. For that, it is convenient to switch to a two-flavor model.}
Although three neutrino flavors appear in nature, a two-neutrino framework serves as a straightforward yet effective approach to capturing the essential aspects of atmospheric neutrino oscillations. When neutrinos travel through matter, their flavor oscillations are modified by the weak interaction as well as torsional four-fermion interaction. The flavor oscillations are controlled by the equation 
%
\begin{equation}
	i\frac{d}{dt}\begin{pmatrix}{\nu_\alpha} \\ {\nu_{\beta}}\end{pmatrix} =\left[\frac{1}{4E}\begin{pmatrix}-\Delta \tilde{m}^2\cos 2\theta + 2EA & \Delta \tilde{m}^2\sin 2\theta\\ \Delta \tilde{m}^2 \sin 2\theta & \Delta \tilde{m}^2 \cos2\theta - 2EA\end{pmatrix}\right]\begin{pmatrix}{\nu_\alpha} \\ {\nu_{\beta}}\end{pmatrix}\,.
	\label{eq:TDSE_for_2nu_2}
\end{equation}
Here $A = \sqrt{2}G_Fn_e$\ is the matter potential due to weak interaction only,  and $\Delta \tilde{m}^2$ is the torsion induced mass-squared difference given by
\begin{align}
\Delta \tilde{m}^2 := \Delta m^2 +2 \tilde{n} E \Delta\lambda\,,
\end{align}
where $\Delta m^2=m_2^2-m_1^2\,$ is the neutrino intrinsic mass squared difference, $\Delta \lambda = \lambda_2 - \lambda_1\,$ is the difference of geometrical coupling constants of neutrinos, and $\tilde{n} = \sum_f \lambda_f n_f = 3\lambda_e n_e\,$ is the contribution from background fermions. 

\subsection{Mikheyev-Smirnov-Wolfenstein resonance}
{The effective Hamiltonian of Eq.~(\ref{eq:TDSE_for_2nu_2}) can not be diagonalized in matter} by the vacuum mixing matrix $U\,.$  Instead, diagonalization occurs by a unitary matrix 
\begin{equation}
U_{\rm eff} = \begin{pmatrix}
\cos \theta_{\rm eff} & \sin \theta_{\rm eff}\\ -\sin\theta_{\rm eff} & \cos\theta_{\rm eff}
\end{pmatrix}\,,
\end{equation}
which is the effective mixing matrix in matter~\cite{Giunti:2007ry}, such that
\begin{align}
U^T_{\rm eff} H^F_{\rm eff} U_{\rm eff} = \frac{1}{4E}\mathrm{diag}(-\Delta m^2_{\rm eff},\Delta m^2_{\rm eff})\,.
\end{align}
 The effective mixing angle and mass squared differences in matter are related to their values in vacuum as~\cite{Barick:2023wxx}
\begin{align}
\sin 2 {\theta}_{\rm eff} = \frac{\Delta \tilde{m}^2 \sin 2 \theta }{\Delta {m}_{\rm eff}^2} \quad , \qquad \Delta {m}_{\rm eff}^2 = \sqrt{ \left(\Delta \tilde{m}^2 \cos 2 \theta - 2EA\right)^2 + \left(\Delta \tilde{m}^2 \sin 2 \theta\right)^2}\,.
\label{mixing-angle-mass-diff-matter}
\end{align}

The effective mixing angle in matter $\theta_{\rm eff}$ will be maximum when $\Delta {m}_{\rm eff}^2 = \Delta \tilde{m}^2 \sin 2 \theta \,,$ and this happens at $\, 2EA = \Delta \tilde{m}^2 \cos 2 \theta \,,$ as can be seen from Eq.~(\ref{mixing-angle-mass-diff-matter}). Thus the required condition for maximum mixing angle in matter is given by                      
\begin{align}
&E = \frac{\Delta m^2 \cos 2 \theta}{\left(2 \sqrt{2}G_F n_e-2\tilde{n}\Delta \lambda \cos2\theta\right)} =: \tilde{E}_{\rm MSW}\,,
\label{E_MSW-with-torsion}
\end{align}
which we can think of as the Mikheyev-Smirnov-Wolfenstein (MSW) resonance  energy~\cite{Mikheyev:1985zog,Wolfenstein:1977ue} in presence of the geometrical interaction.
If we put $\Delta \lambda=0\,$ or $\lambda_e=0\,$ in the above equation, we get the MSW resonance energy in presence 
of weak interaction only, which is discussed extensively in the literature~\cite{Wolfenstein:1977ue,Akhmedov:2006hb,Gandhi:2004md,Gandhi:2004bj,Gandhi:2007td,Mikheyev:1985zog,Kelly:2021jfs}.
In the 2-flavor scenario, the neutrino conversion probability in matter is given by
\begin{align}
P^{2F}_{\alpha \beta} = \sin ^2 2 \theta_{\rm eff} \sin^2 \frac{\Delta {m}_{\rm eff}^2 L} { 4E}. \qquad (\alpha \neq \beta)
\label{prob-2-nu}
\end{align}
Therefore the $ \nu_\alpha \to \nu_\beta$ conversion probability will be maximum when both $\sin^2 2 \theta_{\rm eff} = 1$ and $\sin^2 \frac{\Delta {m}_{\rm eff}^2 L} { 4E} = 1 $. The first condition gives the energy where MSW resonance takes place and from the second one we get
\begin{align}
\frac{\Delta {m}_{\rm eff}^2 L} { 4E} = \frac{(2k+1) \pi}{2} \,, \qquad k=0,1,2,..
\label{msw-baseline}
\end{align} 
{a relation between the neutrino energy and the baseline}. If we put $E = \tilde{E}_{\rm MSW}\,$ in Eq.(\ref{msw-baseline}), we get the corresponding baseline at which $P^{2F}_{\alpha \beta}$ is maximum.  

\begin{figure}[bhtp]
\includegraphics[width=10cm]{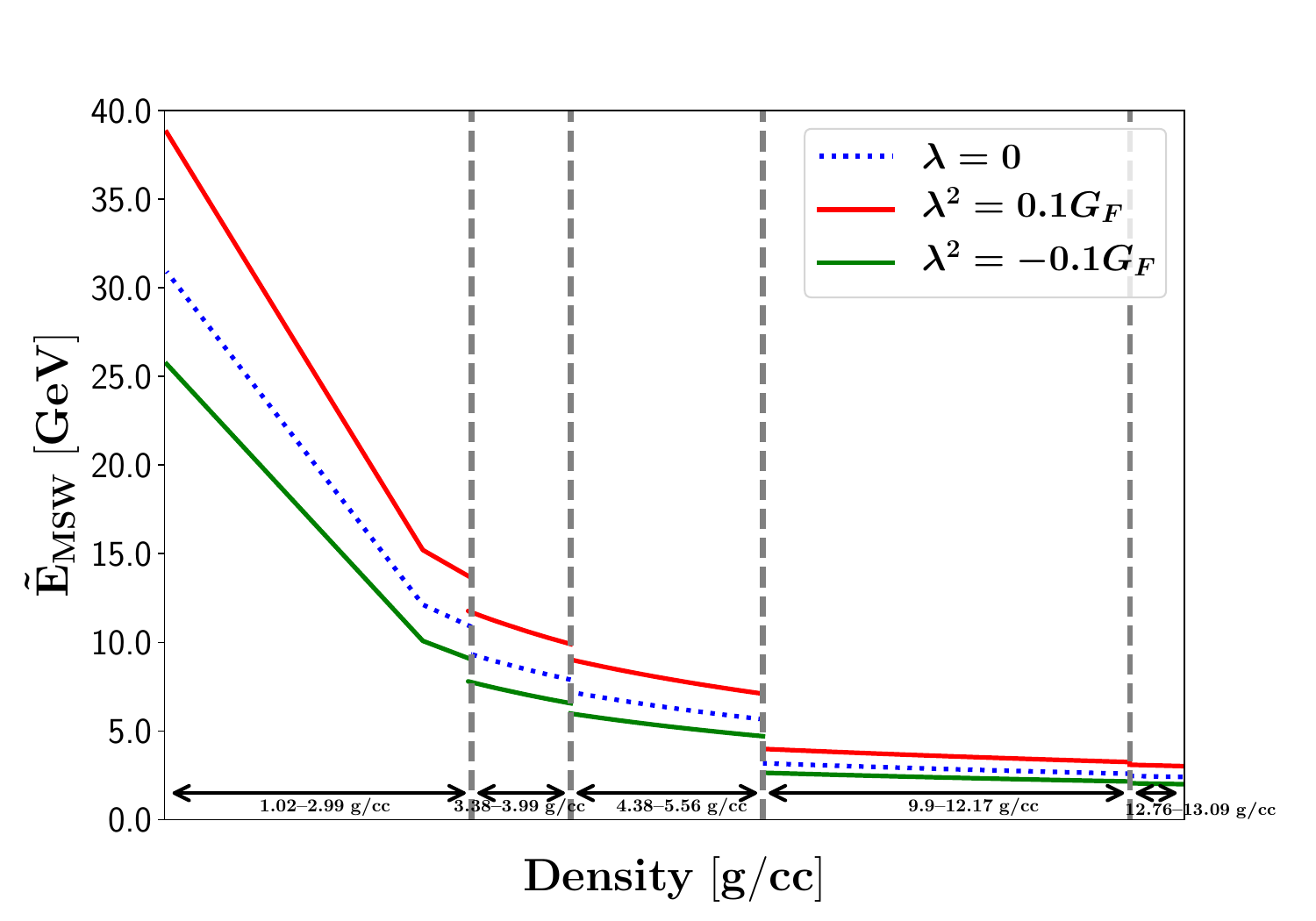}
\caption{MSW resonance energy as a function of the Earth's density, for different $\lambda\,.$}
\label{MSW-rho}
\end{figure}

In order to get a quantitative idea about the MSW resonance energy and baseline, 
{we take $\Delta m^2=2.507\times 10^{-3} \mathrm{eV}^2$ and mixing angle $\theta=8.58^\circ\,,$ relevant for the atmospheric neutrino oscillations~\cite{ParticleDataGroup:2024cfk}, 
the average Earth matter density $\rho=5.5$ gm/cc,
and the geometrical coupling parameter $\lambda^2 =0.1~ G_F\,.$} Now by putting all these values in Eq.~(\ref{E_MSW-with-torsion}), we get the MSW resonance energy in presence of geometrical interaction  as $\tilde{E}_{\rm MSW}=7.2$ GeV. 
{We have calculated {the baseline} according to 
Eq.~(\ref{msw-baseline}) {for both $\lambda^2=0$ and $\lambda^2\neq0\,.$} The  only 
allowed value is $k=0\,,$  because for $k\geq 1\,,$ $L$ exceeds the diameter of the Earth for $E = \tilde{E}_{\rm MSW}$\,.} The minimum baseline required for this resonance to occur is $ L = 9583 $ km, which we get by putting all the values including the value of MSW resonance energy $\tilde{E}_{\rm MSW}\,,$ in Eq.~(\ref{msw-baseline}).{ The corresponding zenith angle for the MSW resonance to occur in presence of torsional four-fermion interaction can be found from Eq.~(\ref{baseline-zenith}), and it comes to $\theta_\nu \sim 139^\circ\,.$} These values of the MSW resonance energy and zenith angle {agree with the oscillograms of Fig.~\ref{oscillogram-e}, Fig.~\ref{oscillogram-mu} and Fig.~\ref{oscillogram-tau}}. The variation of the MSW resonance energy with Earth's matter density,{ according to the PREM profile \cite{Dziewonski:1981xy},} for $\lambda^2=0\, \mathrm{(SI)}\, \mathrm{and}\,\pm 0.1 ~{G_F}$, is shown in Fig.~\ref{MSW-rho}.  It is clear from the plot that the MSW resonance energy increases for $\lambda^2 = 0.1~ G_F$ and decreases for $\lambda^2 = -0.1~ G_F$ compared to the standard scenario $(\lambda = 0)\,$. 

\subsection{Parametric resonance} 
The propagation of neutrinos through a medium with varying density layers has been widely explored in the literature~\cite{Akhmedov:2006hb,Akhmedov:1998ui,Akhmedov:1998xq,Chizhov:1998ug,Chizhov:1999he,Liu:1998nb,Chizhov:1999az,Akhmedov:2005yj}. Here we briefly outline the conditions for parametric enhancement in atmospheric neutrino oscillations. To begin, consider the evolution of a two-flavor neutrino state $\nu = (\nu_\alpha, \nu_\beta)^T\,,$ in the presence of matter. {The evolution operator governing the transition from an initial state $\nu(L_0)$ to a final state $\nu(L)$ is $ S(L-L_0)$\,, which for a time-invariant Hamiltonian is given by $S(L - L_0) = \exp(-iH(L-L_0)\,.$} To analyze parametric resonance, we consider a simplified Earth model with two constant-density layers: the core, with density $\rho_C$ and the mantle, with density $\rho_M$.

The evolution operator for a layer $X (=M,C)$ can always be written in terms of Pauli matrices as
\begin{align}
S_X = \cos \Phi_X \mathbb{I}- i\sin \Phi_X (\vec{\sigma}\cdot\vec{n}_X)\,,
\label{evolution_operator}
\end{align}
with $\Phi_X$ being the oscillation phase in the layer $X$\,,
\begin{align}
\Phi_X = \frac{\Delta m_X^2 L_X}{4E}\,,
\end{align}
$L_X$ the distance travelled within the layer $X$\,, and $\Delta m_X^2$ the total mass squared difference in the layer X, given in Eq.~(\ref{mixing-angle-mass-diff-matter}) as $\Delta {m}_{\rm eff}^2\,.$ {In this section} we have replaced the subscript `${\rm eff}$' by the subscript `$X$' for the different layers. The unit vector $\vec{n}_X$ within the layer $X$ is given by 
\begin{align}
\vec{n}_X = \left( \sin2\theta_X,0,-\cos2\theta_X \right) \,,
\end{align}
where $\theta_X$ is the total mixing angle within the layer X as given in Eq.~(\ref{mixing-angle-mass-diff-matter}).

The neutrinos first move through the mantle then the core and finally again through the mantle. {We can then construct the evolution matrix as the product}
\begin{align}
S = S_M S_C S_M\,.
\end{align}
{This can be written in the form}
\begin{align}
S = W_0 \mathbb{I} - i \vec{\sigma}\cdot\vec{W}\,,
\label{evolution_operator_final_2nu_2layer}
\end{align}
where the expressions for $W_0$ and $\vec{W}=(W_1,0,W_3)$ are given by~\cite{Kelly:2021jfs}
\begin{align}
W_0 &= \cos2\Phi_M \cos\Phi_C - \cos2(\theta_M-\theta_C)\sin2\Phi_M\sin\Phi_C\,, \\
W_1 &= \sin2\theta_M \sin2\Phi_M \cos\Phi_C  \notag\\
& \qquad + \left[\cos2\left(\theta_M-\theta_C\right)\cos2\Phi_M \sin2\theta_M - \cos2\theta_M \sin2(\theta_M-\theta_C)\right]\sin\Phi_C\,, \\ 
W_3 &= -\sin2\theta_M \sin\Phi_C \sin2(\theta_M-\theta_C) \notag \\
& \qquad - \left( \sin2\Phi_M \cos\phi_C + \cos2(\theta_M-\theta_C) \cos2\Phi_M \sin\Phi_C \right)\cos2\theta_M \,.
\end{align}
As expected, $W_0$ and $\vec{W}$ depend on the oscillation phases and mixing angles of both layers. Then the explicit form of the evolution matrix for the propagation of neutrinos through the two different Earth density layers is given by
\begin{align}
S=
\begin{pmatrix}
W_0-iW_3 & -iW_1 \\
-iW_1 & W_0-iW_3
\end{pmatrix}\,.
\end{align}
From the above expression, the survival probability and the conversion probability 
in the two-neutrino scenario are given by 
\begin{align}
P_{\alpha \alpha} = |W_0|^2+|W_3|^2 = P_{\beta \beta} \,, \qquad P_{\alpha \beta} = |W_1|^2\,.
\end{align}
Complete flavor conversion occurs when $P_{\alpha \alpha}=P_{\beta \beta} = 0\,,$   i.e., when
\begin{align}
W_0 =0 \qquad \mathrm{and} \qquad W_3=0\,.
\end{align}
The solution of the above set of equations is given by
\begin{align}
\tan^2\Phi_M = -\frac{\cos2\theta_C}{\cos2(2\theta_M-\theta_C)} \qquad \mathrm{and} \qquad \tan^2\Phi_C = -\frac{\cos^2(2\theta_M)}{\cos2\theta_C \cos2(2\theta_M-\theta_C)}\,.
\label{parametric-reso-baseline}
\end{align}
Thus the oscillation phase in a layer depends on the mixing angle in both layers. A parametric resonance occurs when
\begin{align}
\cos2\theta_C \leq 0 \quad &\mathrm{and} \quad \cos2(2\theta_M-\theta_C) \geq 0\,, \notag \\
\mathrm{or} \qquad & \notag \\
\cos2\theta_C \geq 0 \quad &\mathrm{and} \quad \cos2(2\theta_M-\theta_C) \leq 0\,.
\label{parametric-resonance-condition}
\end{align}
{The first condition is satisfied only for $\Delta m^2 > 0\,,$ whereas the second condition is satisfied only for $\Delta m^2 < 0\,.$}
So the condition for parametric resonance depends on the total mixing angle in each layer $\theta_M$ and $\theta_C$ and thus the density of each layer $\rho_M$ and $\rho_C$ and the energy of the incoming neutrino $E$. For example, if we consider an average mantle density of $4.5$ gm/cc and and average core density of $11.5$ gm/cc, then from Eq.~(\ref{parametric-resonance-condition}), we find that the allowed energy range for parametric resonance to occur with $\lambda^2 = 0.1~G_F\,,$ is 3.5~{\rm GeV}~$ < E < $~11.6~{\rm GeV}, {which agree with the oscillograms of Fig.~\ref{oscillogram-e}, Fig.~\ref{oscillogram-mu} and Fig.~\ref{oscillogram-tau}}.

In regions where parametric resonance can occur, the neutrino path lengths needed for complete flavor conversion can be inferred from the solutions given in Eq.~(\ref{parametric-reso-baseline}). It is important to note however, that for atmospheric neutrinos, the distances traveled through different Earth layers are not independent, rather, they are all related to the zenith angle at which neutrinos cross the Earth. In a simplified two-layer model of the Earth, the baselines $L_M$ through the mantle  and $L_C$ through the core  can be written in terms of the zenith angle $\theta_\nu$ as 
\begin{equation}
L_C = 2R\sqrt{f^2_c-\sin^2\theta_\nu} \qquad \mathrm{and} \qquad L_M = -2R\left(\cos\theta_\nu + \sqrt{f^2_c-\sin^2\theta_\nu}\right)
\label{baseline-mantle-core}
\end{equation}
where $f_c \equiv \frac{R_C}{R}\,,$ $R_C$ being the core radius and $R$ the radius of the Earth.

Thus for {given mantle and core densities}, one can determine the combinations of neutrino energy and zenith angle that give rise to parametric     resonance. However these combinations are not unique, multiple parameter configurations may lead to resonant behavior. 
{For example, for the $\rho_M$ and $\rho_C$ considered above and neutrino energy $E = 4 $ GeV, the zenith angle at which full flavor conversion occurs  for $\lambda^2 = 0.1~G_F\,$ is $\theta_\nu = 150.8^\circ\,,$ which agrees with the oscillograms given in section~\ref{3nu}. However, a different set of values of $\rho_M$\,, $\rho_C$\,, and $E$ would lead to a different zenith angle for full flavor conversion for the same $\lambda^2$.  }



\subsection{Resonance for three flavors}
So far we have discussed the MSW and parametric resonance condition in a two flavor scenario. In this section we will extend our discussion for a more realistic three flavor case. Our discussion will be based on the approximations of constant matter density and one mass scale dominance (OMSD) which neglects the smaller mass squared difference $\Delta m^2_{21}$ compared to $\Delta m^2_{31}$~\cite{Gandhi:2004md, Gandhi:2004bj, Gandhi:2007td}. {Additionally, let us also neglect $\Delta \lambda_{21}$. In this case,} the oscillation probabilities become independent of the mixing angle $\theta_{12}$ and the CP phase $\delta_{CP}$. This approximation allows for simplified analytical expressions, making it easier to qualitatively interpret matter effects.

 In vacuum, the $\nu_\mu \to \nu_e$ conversion probability is 
 \begin{equation}
     P^{\rm vac}_{\mu e} = \sin^2\theta_{23} \sin^2 2 \theta_{13} \sin^2 \left(\frac{\Delta m^2_{31}L}{4E}\right)\,.
\label{mu-e-prob-vac}
 \end{equation}
Under the constant-density approximation, matter effects can be incorporated by substituting $\theta_{13}$ and $\Delta m^2_{31}$ with their effective values in matter, $\theta_{\rm eff}$ and $\Delta m^2_{\rm eff}$, as given in Eq.~(\ref{mixing-angle-mass-diff-matter}). Thus $\nu_\mu \to \nu_e$ conversion probability in matter is given by
 \begin{equation}
     P^{\rm mat}_{\mu e} = \sin^2\theta_{23} \sin^2 2 \theta_{\rm eff} \sin^2\left(\frac{\Delta m^2_{\rm eff}L}{4E}\right) = \sin^2\theta_{23} P^{2F}_{\alpha \beta}\,,
 \end{equation}
where $P^{2F}_{\alpha \beta}$ is the 2-flavor conversion probability as calculated above in Eq.~(\ref{prob-2-nu}). {Thus, in the three-neutrino case, the conditions for MSW and parametric resonances remain the same, the only difference is that the probability expression includes an additional factor of $\sin^2\theta_{23}$.}

\section{Summary and outlook \label{summary}}
In this paper, we have focused on the effect of the geometrical (torsional) four-fermion interaction on atmospheric neutrino oscillations. The spacetime torsion here is not a background spacetime field, but is dynamically generated by the fermions. When neutrinos propagate through matter, they experience potential due to weak interactions as well as the four-fermion interaction arising from eliminating the spacetime torsion. The first one is diagonal in the flavor basis, whereas the second one is diagonal in the mass basis. By including all the effects coming from both of these interactions, we have numerically solved the Schr\"odinger equation for three flavors of neutrinos, using the PREM density profile of the Earth. {We have plotted oscillograms for $P_{\mu e}\,, P_{\mu \tau}$ and $P_{\mu \mu}\,,$ for $\lambda=0$ (SI) and $\lambda^2=\pm 0.1~G_F\,,$ for both NH and IH}. {For the normal hierarchy (NH), both MSW and parametric resonances appear in the presence of geometrical interaction (\(\lambda^2 = \pm 0.1~G_F\)), similar to the standard interaction scenario. The $\nu_\mu \to \nu_e\,,\,\nu_\mu \to \nu_\tau$  conversion probabilities and the $\nu_\mu$ survival probability show significant deviations from the \(\lambda = 0\) case, especially in the {core region}. A visible change in the high-energy region occurs only for \(\lambda^2 > 0\), because the contribution $2\tilde{n}\Delta \lambda E \equiv 2n\lambda^2E$ due to the torsional four-fermion interaction adds to $\Delta m^2$ in the oscillation probability expressions. For the {inverted hierarchy (IH)}, no MSW or parametric resonances are observed for $\lambda^2 = 0$ or $\pm 0.1\,G_F$, although oscillation probabilities still change when $\lambda \neq 0$.
}

Since the geometrical interaction contributes to the off-diagonal terms in the {effective} mixing matrix, its presence is expected to influence the dependence of conversion probabilities on $\delta_{CP}\,.$ To explore this effect, we have analyzed the dependence of $P_{\mu \tau}$ on $\delta_{CP}$ in the presence of torsional four fermion interaction. In the SI case, $P_{\mu \tau}$ exhibits only a weak dependence on $\delta_{CP}$ in the low-energy region, and for energies above $15$ GeV, it becomes entirely independent of $\delta_{CP}$. However for $\lambda^2 = 0.1 ~ G_F\,,$ $P_{\mu \tau}$ retains its dependence on $\delta_{CP}$ even at very high energies, around $E\sim200$ GeV. 
{We have also investigated how the MSW and parametric resonances are affected by the geometrical interaction. We have calculated the MSW and parametric resonance energy and the corresponding baseline or the zenith angle, after including the effect of geometrical interaction, for the two-neutrino scenario. Extending the result for the three-neutrino case and using OMSD approximation, we can match the results with the oscillograms. } 

{The main purpose of this paper was to get an idea about how the geometrical interaction affects flavor oscillations of atmospheric neutrinos. Event-level calculations are needed to derive constraints on the geometrical coupling parameters using {simulated data} for upcoming atmospheric neutrino experiments {as well as actual data received from existing experiments}. Further, a detailed $\chi^2$-analysis is needed to assess the impact of this novel interaction on key oscillation physics sensitivities —-- specifically, CP violation, the $\theta_{23}$ octant, and the neutrino mass hierarchy. We plan to report on these questions in future work.}

{\bf Acknowledgments:} R.B. would like to thank Sharmistha Chattopadhyay and Srubabati Goswami for fruitful discussions.

\appendix
\section{Analytical expressions}\label{analytics}
{In this Appendix, we give the analytical expressions for various probabilities of flavor conversion and survival following from Eq.~(\ref{eq:TDSE_for_3nu.b}). In the approximation of small $\tilde{\alpha}$ and $\sin\theta_{13}\,,$ we keep terms up to second order in them to find these probabilities. For more details,  we refer the reader to~\cite{Barick:2023wxx,Barick:2023qjq}.}

{For the the conversion of muon neutrinos into other flavors, 
the $\nu_\mu \to \nu_\tau$ transition probability can be expressed as (with $E_{ij} = E_i - E_j$)
\begin{align}
P_{\mu\tau}= &2s_{23}^2c_{23}^2(1-\cos(E_{23}L))-2\frac{\tilde{\alpha}^2}{\tilde{A}^2}s_{12}^2c_{12}^2s_{23}^2c_{23}^2(1-\cos(E_{12}L)+\cos(E_{13}L)-\cos(E_{23}L)) \nonumber \\
&-2\frac{s_{13}^2}{(\tilde{A}-1)^2}s_{23}^2c_{23}^2(1+\cos(E_{12}L)-\cos(E_{13}L)-\cos(E_{23}L)) \nonumber \\
&+2\tilde{\alpha} s_{13}s_{12}c_{12}s_{23}^3c_{23}\left[\left(1+\frac{1}{\tilde{A}}\right)(\cos\delta-\cos(E_{23}L+\delta))\right. \nonumber \\
&\left.+\frac{1}{\tilde{A}(\tilde{A}-1)}(\cos(E_{12}L+\delta)-\cos(E_{13}L+\delta))+\frac{\tilde{A}}{\tilde{A}-1}(\cos\delta-\cos(E_{23}L-\delta))\right] \nonumber \\
&+2\tilde{\alpha} s_{13}s_{12}c_{12}s_{23}c_{23}^3\left[\left(1+\frac{1}{\tilde{A}}\right)(-\cos\delta+\cos(E_{23}L-\delta))\right. \nonumber \\
&\left.+\frac{1}{\tilde{A}(\tilde{A}-1)}(-\cos(E_{12}L-\delta)+\cos(E_{13}L-\delta))+\frac{\tilde{A}}{\tilde{A}-1}(-\cos\delta+\cos(E_{23}L+\delta))\right]\,.
\label{mutau1}
\end{align}
Similarly, the $\nu_\mu \to \nu_e$ conversion probability can be written as
\begin{align}
P_{\mu e}= &\frac{\tilde{\alpha}^2}{2\tilde{A}^2} \sin^2(2\theta_{12}){c}_{23}^2 \left( 1-\cos({E_{12}L}) \right) +2 \frac{{s}_{13}^2}{(\tilde{A}-1)^2} {s}_{23}^{2}\left( 1-\cos({E_{13}L}) \right) \nonumber \\
	&+\frac{1}{\tilde{A}(\tilde{A}-1)}2\tilde{\alpha} {s}_{13}s_{12}c_{12}s_{23}c_{23}(\cos\delta - \cos(E_{12}L+\delta) - \cos(E_{13}L-\delta) + \cos(E_{23}L-\delta)) \,.	
	 \label{eq:nu_mu_to_nu_e}
\end{align}
The $\nu_\mu$ survival probability is thus 
\begin{align}
P_{\mu \mu}= &1-2c_{23}^2s_{23}^2(1-\cos(E_{23}L)) \nonumber \\
&+2\frac{\tilde{\alpha}^2}{\tilde{A}^2}s_{12}^2c_{12}^2c_{23}^2\left[c_{23}^2\cos(E_{12}L)+s_{23}^2\cos(E_{13}L)-c_{23}^2-s_{23}^2\cos(E_{23}L)\right]\nonumber \\
&+2\frac{s_{13}^2}{(\tilde{A}-1)^2}s_{23}^2\left[c_{23}^2\cos(E_{12}L)+s_{23}^2\cos(E_{13}L)-c_{23}^2\cos(E_{23}L)-s_{23}^2\right] \nonumber \\
&+4\tilde{\alpha} s_{13}s_{12}c_{12}s_{23}c_{23}\cos\delta\left[c_{23}^2\frac{1}{\tilde{A}(\tilde{A}-1)}\cos(E_{12}L)+s_{23}^2\frac{1}{\tilde{A}(\tilde{A}-1)}\cos(E_{13}L)\right. \nonumber \\
&\left.+\left(1+\frac{1}{\tilde{A}}\right)c_{23}^2-c_{23}^2\frac{\tilde{A}}{\tilde{A}-1}\cos(E_{23}L)+s_{23}^2\left(1+\frac{1}{\tilde{A}} \right)\cos(E_{23}L)-\frac{\tilde{A}}{\tilde{A}-1}s_{23}^2\right]\,.
\label{eq:nu_mu_to_nu_mu}
\end{align}
Our results are consistent with those in ~\cite{Akhmedov:2004ny, Nunokawa:2005nx, Barger:1980tf, Zaglauer:1988gz, Yasuda:1998mh, Kimura:2002wd, Freund:2001pn} within the energy range $0.5~\text{GeV} < E < 10~\text{GeV}$, when the geometrical couplings are set to zero.}


\bibliographystyle{unsrt}
\bibliography{references}

\end{document}